\documentclass[useAMS,usenatbib]{mn2e}
\usepackage{longtable}
\usepackage{lscape}
\usepackage{graphicx}

\newcommand{\Msun}{\mbox{M$_{\odot}$}}
\title[The sub-stellar birth rate from UKIDSS]{The sub-stellar birth rate from UKIDSS}
\author[A. C. Day-Jones et al.]{A. C. Day-Jones$^{1,2}$\thanks{E-mail:adjones@das.uchile.cl. Based on observations made with ESO telescopes at the La Silla Paranal Observatory under programmes 086.C-0450, 087.C-0639 \& 088.C-0048.}, F. Marocco$^2$, D. J. Pinfield$^2$, Z.H. Zhang$^2$, B. Burningham$^2$,\\ \newauthor N. Deacon$^4$, M.T. Ruiz$^1$, J. Gallardo$^5$, H.R.A. Jones$^2$, P.W.L. Lucas$^2$,J.S. Jenkins$^1$, \\ \newauthor  J. Gomes$^2$, S.L. Folkes$^{3,2}$, J.R.A. Clarke$^{3,2}$.\\$^{1}$ Departamento de Astronomia, Universidad de Chile, Camino del Observatorio 1515, Santiago, Chile. \\$^{2}$  Centre for Astrophysics Research, University of Hertfordshire, College Lane, Hatfield, Hertfordshire, UK.\\$^{3}$ Departamento de Fisica y Astronomia, Facultad de Ciencias, Universidad de Valparaiso, Av.Gran Bretana 1111, Valparaiso, Chile.\\$^4$ Max Planck Institute for Astronomy, Königstuhl, 17. D-69117, Heidelberg, Germany.\\$^5$ ALMA, Alonso de Cordova 3107, Vitacura, Santiago, Chile.
}
\begin{document}

\date{}

\pagerange{\pageref{firstpage}--\pageref{lastpage}} \pubyear{2012}

\maketitle

\label{firstpage}

\begin{abstract}
We present a new sample of mid L to mid T dwarfs with effective temperatures of 1100 to 1700 K selected from the UKIDSS Large Area Survey and confirmed with infrared spectra from X-Shooter/VLT. This effective temperature range is especially sensitive to the formation history of Galactic brown dwarfs and allows us to constrain the form of the sub-stellar birth rate, with sensitivity to differentiate between a flat (stellar like)
birth rate, and an exponentially declining form.  We present the discovery of 63 new L and T dwarfs from the UKIDSS LAS DR7, including the identification of 12 likely unresolved binaries, which form the first complete sub-set from our program, covering 495 sq degrees of sky, complete to J=18.1. We compare our results for this sub-sample with simulations of differing birth rates for objects of mass 0.10-0.03M$_\odot$ and ages 1-10Gyrs. We find that the more extreme birth rates (e.g. a halo type form) can likely be excluded as the true form of the birth rate. In addition we find that although there is substantial scatter we find a preference for a mass function, with a power-law index, $\alpha$ in the range ${-}1 < \alpha < 0$ that is consistent (within the errors) with the studies of late T dwarfs.
\end{abstract}


\section{Introduction}
\label{intro}
The distribution of star formation with mass and time are key pieces of observational evidence for understanding star formation in the galaxy. The former is described by the initial mass function (IMF; \citealt{salpeter55}), which can be described as a power-law of the form $\psi(M)~\propto~M^{-\alpha}$, with $\alpha = 2.35$, and has been determined across the stellar mass regime by measuring the luminosity function for a population of stars, and applying a mass-luminosity relation, which should account for metallicity variations. Since brown dwarfs never reach the main sequence, this determination is complicated in the sub-stellar regime by the lack of a unique mass-luminosity relationship. Instead the $T_{\rm eff}$ and luminosity are dependent on mass \emph{and} age (\citealt{allard97}). This means that the luminosity function and $T_{\rm eff}$ distributions of field brown dwarfs depend not only on the mass function, but also on their formation history (\citealt{chabrier02}). Indeed, extending the Salpeter mass function to sub-stellar objects one would expect many more brown dwarfs than stars, which is not seen by observations of late M and L dwarfs (e.g. \citealt{reid02}). In addition it is not totally apparent that the full field population of brown dwarfs is populated by objects that formed through a cloud fragmentation process. There are a number of ways that brown dwarfs could form that are different to the canonical ways in which stars are thought to form. There are at least four different formation mechanisms that have been suggested, such as formation through the ``ejection'' of pre-stellar cores (\citealt{reipurth01}; \citealt{bate02}; \citealt{delgado03}; \citealt{sterzik03}) or through ``turbulence'', or turbulent fragmentation (\citealt{padoan02}; \citealt{padoan04}). Other theories include that of ``disc fragmentation'', forming sub-stellar cores from an initially massive pre-stellar core via fragmentation
of a large circumstellar disk (\citealt{boffin98}; \citealt{bate03}; \citealt{whitworth05}; \citealt{whitworth06}) and ``photo-erosion'', where sub-stellar objects form in the presence of a higher mass star embedded in a HII region (\citealt{whitworth04}). As the formation mechanisms could indeed be different for sub-stellar objects it is thus important to define the mass function and formation history in the sub-stellar regime if we wish to fully understand their contribution to the Galactic population.

Young clusters have been the target of many studies seeking to measure the sub-stellar IMF since their known ages and metallicities allow the use of a mass-luminosity relation based on a single coeval age (e.g. \citealt{caballero09}; \citealt{oliveira09}; \citealt{luhman07}; \citealt{lodieu07a}). Although these clusters allow a relatively direct measurement of the sub-stellar IMF, they also introduce their own problems, since the initial conditions and accretion histories of individual objects introduce uncertainties regarding the ages, and hence masses, of such young objects (e.g. \citealt{baraffe10}). 
As it is difficult to determine the age of field brown dwarfs, unless they have fiducial constraints on their age as binaries (e.g. \citealt{burningham09}; \citealt{zhang10}; \citealt{burningham10a}; \citealt{dayjones11}; \citealt{murray11}) or as members of moving groups (\citealt{ga10}; \citealt{clarke10}), estimating the mass function of field brown dwarfs requires a knowledge of their formation history.  This is often assumed to be the same as that for stars (constant with time; \citealt{miller79}), but is unconstrained in the sub-stellar regime. 
The first attempt to measure the sub-stellar mass function was made by \citet{reid99} from preliminary results from 2MASS data based on only 17 L dwarfs. More recently, with the discovery of late T dwarfs several other groups have made measurements of the sub-stellar mass function in the disc. These have all generally been with small sample sizes or cover only L dwarfs (e.g. \citealt{cruz07}) or only T dwarfs (e.g. \citealt{burningham10b}; \citealt{kp12}; \citealt{metchev08}). Those that have considered the full temperature regime across the L and T dwarf spectral types (e.g \citealt{reyle10}) suffer from large associated errors and large bin sizes in order to get large enough sampling.
 
In order to characterise the form of the sub-stellar formation history a large sample of brown dwarfs is required. With modern large-scale near and mid infrared surveys, such as the 2MASS (\citealt{sk06}), UKIDSS (\citealt{lawrence07}), VISTA (\citealt{emerson03}) and WISE (\citealt{wright10}), which have identified large numbers of brown dwarfs (\citealt{kp00}; \citealt{hawley02}; \citealt{pinfield08}; \citealt{burningham10b}; \citealt{kp11}, \citealt{cushing11}) it is now possible to provide the necessary sample of such objects.  This paper outlines our efforts to use the UKIDSS Large Area Survey to empirically constrain the Galactic brown dwarf formation history. We discuss past and present simulations of the formation history in $\S$\ref{form}. In $\S$\ref{sample} the selection of our sample of L and T dwarfs from the UKIDSS LAS, $\S$\ref{obs} gives details of the observations and data reduction. In $\S$\ref{spec} we present the spectroscopy and spectral types of our sample and investigate potential unresolved binarity. In $\S$\ref{fh} we compare our first observations with simulations and look at constraints that can be placed on the formation history of Galactic brown dwarfs. Finally we summarize our findings in $\S$ \ref{concs}.

\section{Simulations of the formation history}
\label{form}
Simulations of the effect of how the birth rate affects the luminosity function have been performed by several authors. Based on model data and direct comparisons with DENIS and 2MASS observations, \citet{chabrier02} made simulations using two different initial mass functions and birth rates. Namely a flat, or constant birth rate, which is the simplest form and supported by the work of \citet{miller79}, who suggest that the birth rate does not depend strongly on the gas density, and is approximately consistent across the Galactic disc. They also consider an exponential form where the formation decreases with time. These are considered with initial mass functions of a power-law form derived by \citet{chabrier01}, and a log normal and exponential form which essentially give the same result when considering the effects from the birth rate.  \citet{burgy04} considered a wider range of birth rates in his monte-carlo simulations considering, in addition to a flat and exponential forms, an ``empirical'' birth rate, which is the same as that measured for stars by \citet{rocha00}, which represents $'$bursts$'$ of formation at peak internavls of 0-1, 2-5 and 7-9Gyrs. This formation history scenario is also supported by the more recent work of the stellar formation history by \citet{cignoni06} and \citet{wyse08}. In addition \citet{burgy04} also considers a ``cluster'' birth rate which assumes a flat, but stochastic (i.e in a number of clusters) formation, which produces a similar result to a flat formation scenario. Finally he considers a ``halo'' type birth rate, that includes formation within a 1~Gyr burst, 9~Gyr in the past, in an attempt to explain a number of subdwarf brown dwarfs that have been identified (e.g. \citealt{burgy03b}; \citealt{lodieu10}).  This scenario gives a radically different $T_{\rm eff}$ distribution for L and T dwarfs compared to the other scenarios, and seems unlikely since we are now seeing a larger number of L dwarfs identified in very young clusters (e.g. Taurus; \citealt{luhman09}, \citealt{quanz10}, Cameleon; \citealt{luhman07b}, Serpens; \citealt{lodieu02}, TWA Hydra; \citealt{chauvin04}, Upper Sco; \citealt{lodieu08}, \citealt{lafren08}) . 

\begin{figure*}
\includegraphics[width=120.0mm,angle=0]{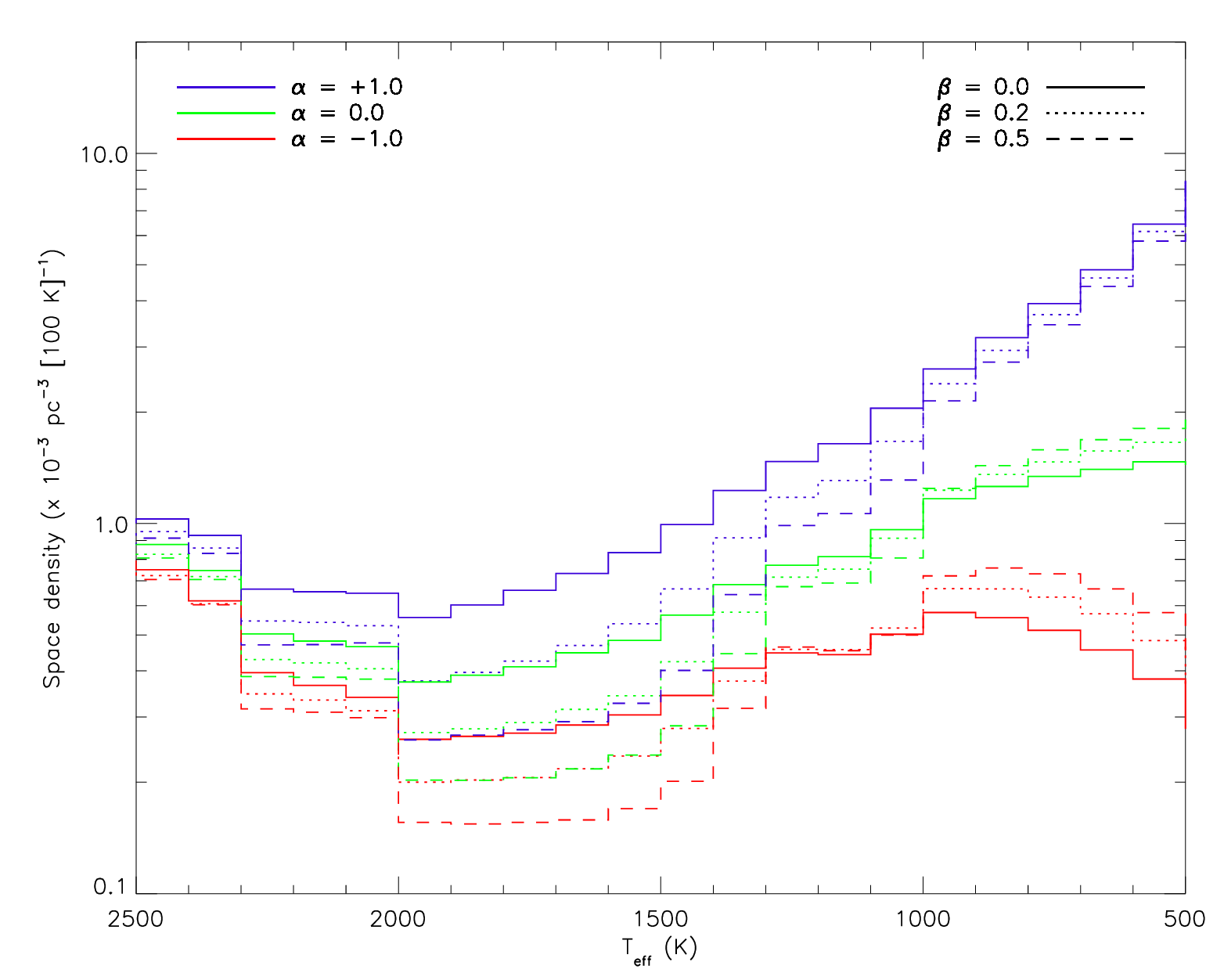}
\caption{Monte-Carlo simulations of differing IMFs, where $\psi(M)~\propto~M^{-\alpha}$ for values of $\alpha = +1.0, 0.0$ and $-1.0$  and birth rates $\beta =0.0, 0.2, 0.5$,  based on the simulations from \citet{deacon2006}.}
\label{niallsimulations}
\end{figure*}

More recent simulations performed by \citet{deacon2006} looked more specifically at L and T dwarfs from the UKIDSS LAS. They produced simulations that take into account several IMFs including a flat, log normal and different power laws ($\alpha =$+1.0, 0, and -1.0), combined with different exponential forms of the birth rate, similarly to those described above. These simulations also included the effect of Galactic disc heating, which had not been included in previous simulations of the birth rate. A histogram of these simulations (a log normal form of the IMF with different birth rates) is shown in Fig.~\ref{niallsimulations}. These simulations are similar to those of \citet{allen05} (see their fig.2). with the main differences arising from the differences in the normalisation of the space density. \citet{allen05} uses 0.35~stars pc$^{-3}$ M$_\odot$ according to \citet{reid97}, where as the simulations based on those of \citet{deacon2006} use 0.0024~stars pc$^{-3}$M$_\odot$ according to \citet{deacon08}. In addition changes also arise from the different values used for their birth rates, and as such are similar and show the same trends but are not directly comparable.

It can be clearly seen in the \citet{deacon2006} simulations that the sub 1000K region is extremely sensitive to the IMF, but relatively insensitive to the birth rate, while the 1100-1500K, corresponding to the  mid-L to mid-T spectral range is most sensitive to differing birth rates. While several hundred brown dwarfs have been identified in large area surveys such as 2MASS and SDSS, they were more sensitive to the detection of L dwarfs and produced only a few tens of early T dwarfs. As such they could not provide the population needed to study and constrain the birth rate. The UKIDSS LAS probes to greater depth across the L and T dwarf spectral types and can provide a statistically robust sample spanning the mid-L to mid-T region, which is most sensitive to the effects of the form of the birth rate. We thus select a sample of mid L -- mid T dwarfs from the UKIDSS LAS in order to compare the space density with that of late T's (\citealt{burningham10a}) and late Ms and earlier Ls to measure the birth rate.

\section{Sample Selection}
\label{sample}

We select our sample from DR7 of the UKIDSS LAS survey, selecting objects with declination $\le$ 20 degrees, $J < 18.1$ and $Y-J>0.8$ following the criteria of \citet{hewett06}, who demonstrate that using a $Y-J > 0.8$ colour criterion is largely free of M dwarfs while including all L dwarfs $\S$\ref{comp}. We also used the following quality flags to ensure that selected objects are point sources and are not likely cross talk effects, or sit at the edge of the detector, such that the following quality flags were used:
\begin{center}
 (priOrSec$ = 0$ OR priOrSec $=$ frameSetID) \\
        yppErrBits  $ < $  256  \\
       j$_{1}$ppErrBits $ < $ 256 \\
       hppErrBits   $ < $ 256 \\
$-3.0 <$ mergedClass $<-0.5$ \\
$-3.0 <$ mergedClassStat $<3.0$ \\
yEll $ < $ 0.45 \\
 j$_{1}$1Ell $ <$  0.45\\
\end{center}
This list of candidates was then cross-matched against SDSS DR7 to identify objects with optical counterparts. Using a matching radius of 4 arcsec we selected objects with optical counterparts according to the criteria below, based on those of \citet{schmidt10}, who provide colours from an unbiased spectroscopically complete sample. Since our NIR colour selection effectively removes all contaminant field M dwarfs this allowed us to be more liberal with our redder sources ($J-K > 1.0$) in terms of our $z-J$ colour selection, and allows for larger uncertainties in the $i'$ band as we are probing the faint end of SDSS. We also required $Y$ and $J$ photometry to have errors not greater than 3$\sigma$ and detections in $H\ge 14.5$, such that this search space would not have been probed by 2MASS. In addition we also consider $K$ band non-detections, if their $z-J$ colour passes our following criteria:  
 
\begin{center}
$J \le 18.1$\\
$Y-J \ge 0.8$\\
 $z - J \ge 2.4$ and ($J - K \ge 1.0$ or no $K$ detection) OR\\
  $z - J \ge 2.9$ and ($J - K < 1.0 $ or no $K$ detection) \\
We then removed objects with the following:\\
 $i-z < 2.0$ and $\sigma (i-z) < 0.35$ \\
 $i-J < 4.7$ and $\sigma(i-j) < 0.2$\\
 $z-K < 3.5$ and $J-K > 1.0$ and $\sigma (z-K) < 0.2$\\
\end{center}

\begin{figure*}
\includegraphics[width=120.0mm,angle=90]{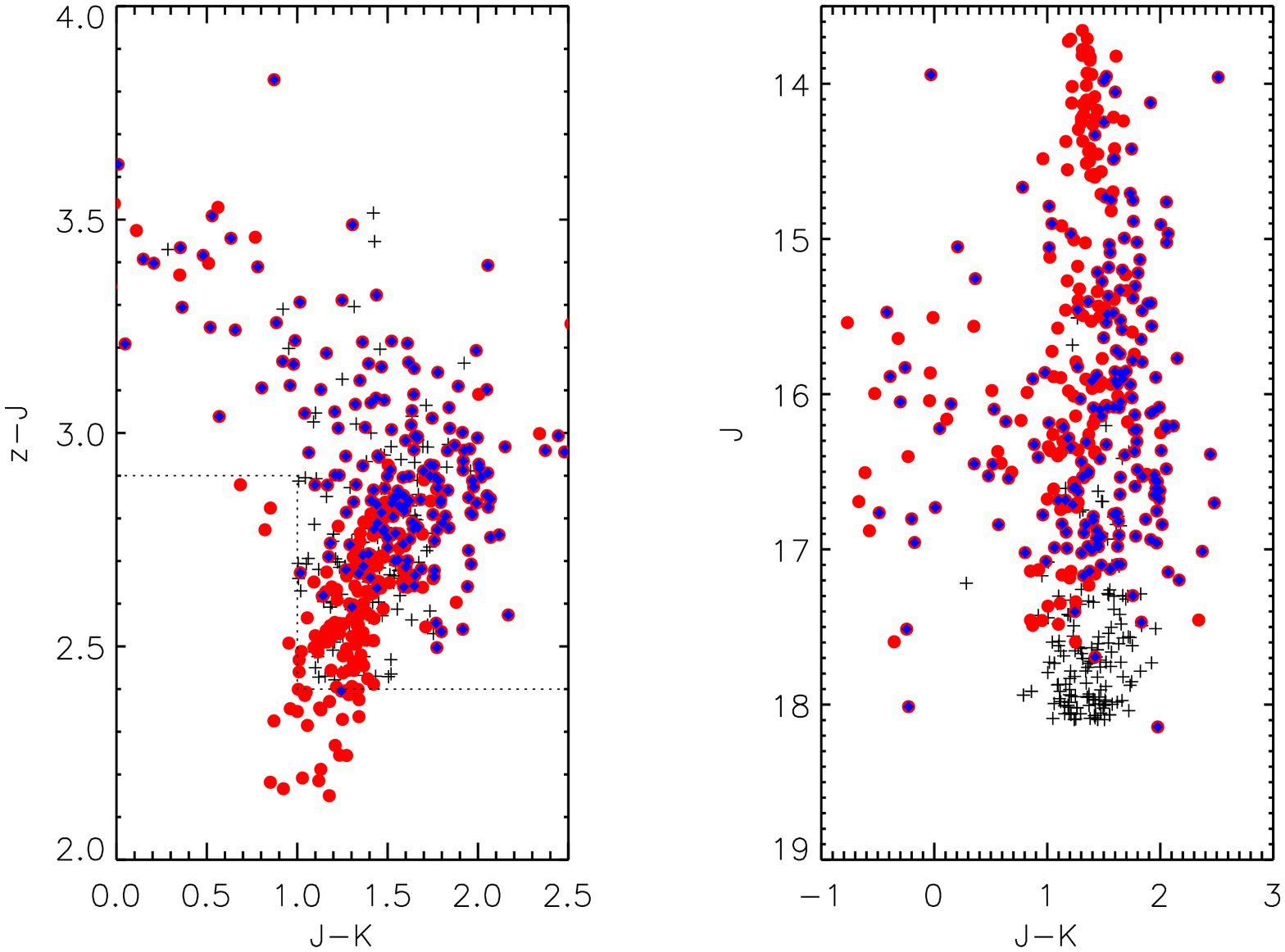}
\caption{Left: Colour-colour diagram of candidate L and T dwarfs (crosses). Right: Colour-Magnitude diagram of candidate L and T dwarfs (crosses). Known L and T dwarfs (L0-T9) from Dwarfarchives.org are shown as filled red circles. Those with spectral types L3-T5 dwarfs are shown as small blue diamonds. Photometry is on the UKIDSS and SDSS system. Objects with 2MASS photometry have been converted using the colour conversions of \citet{warren06}.}
\label{samplecolors}
\end{figure*}

As mid-T dwarfs typically have $z'-J > 3.0$ (e.g.\citealt{pinfield08}) some objects will be too faint for detection in SDSS, we thus include these SDSS non-detections. All objects were then visually inspected to remove any possible miss matches or cross-talk. We also cross matched our sample with known L and T dwarfs in dwarfarchives and retain those that have spectral types of $\ge$L4 (to be consistent with our completeness, see $\S$\ref{comp}), giving a total sample size of 324 from 2000deg$^2$ of UKIDSS LAS (DR7) sky. In this work we consider a sub-sample from a smaller are of 495 sq degrees of sky in the RA and DEC range, RA=22 to 4~hr and DEC=-2 to 16~deg. This includes 76 L and T dwarfs, 13 of which were previously identified (and spectrally typed as L4-T4), 5 of these have been re-observed and 63 new L and T dwarfs. We show photometric information of this sub-sample in Fig.~\ref{samplecolors} and Table.~\ref{photom}, which are discussed and analysed further in the following sections.

\begin{table*}
\caption{Photometry of sub-sample.} 
\centering
\begin{tabular}{|l|l|c|c|c|c|c|c|c|c|}
\hline
Name & ID & RA & DEC & J$^1$ & Y-J$^1$& J-H$^1$ & J-K$^1$ & z-J$^{1,2}$ & i-z$^2$ \\
\hline
ULAS J0006+1540   &BRLT1 & 00:06:13 & +15:40:21 & 17.88 & 1.08 & 1.16 & 1.73 & 2.73 & 3.55 \\
ULAS J0010+0100	& BRLT2 & 00:10:41 & +01:00:13 & 18.09 & 1.27 & 0.63 & 1.38 & 2.71 & 1.42 \\
ULAS J0018-0025	& BRLT3 & 00:18:37 & -00:25:59 & 17.67 & 1.06 & 1.06 & - & 2.61 & 2.83 \\
ULAS J0024+1347	& BRLT6 & 00:24:06 & +13:47:05 & 18.02 & 1.26 & 0.69 & 1.50 & 6.04 & 0.62 \\
ULAS J0027+1423	& BRLT7 & 00:27:07 & +14:23:49 & 17.98 & 0.98 & 0.61 & 1.12 & 2.43 & 2.23 \\
ULAS J0028+1423	& BRLT8 & 00:28:28 & +14:23:49 & 17.56 & 1.38 & 0.95 & 1.72 & 2.72 & 2.50 \\
ULAS J0029+1456	& BRLT9 & 00:29:12 & +14:56:05 & 17.56 & 1.25 & 0.64 & 1.23 & 2.41 & 2.14 \\
SDSSp J0033+1410$^{a}$	& BRLT10 & 00:33:00 & +14:10:37 & 16.65 & 1.18 & 0.96 & 1.64 & 2.78 & 3.66 \\
ULAS J0037-0054	& BRLT12 & 00:37:16 & -00:54:05 & 18.09 & 1.36 & 0.76 & 1.42 & 2.71 & 1.79 \\
ULAS J0043+1411	& BRLT14 & 00:43:56 & +14:11:18 & 17.33 & 1.09 & 0.64 & 1.21 & 2.42 & 2.05 \\
ULAS J0047+1546	& BRLT15 & 00:47:57 & +15:46:41 & 17.83 & 1.29 & 0.66 & 1.41 & - & - \\
ULAS J0050-0003	& BRLT16 & 00:50:38 & -00:03:37 & 17.86 & 1.17 & 0.79 & 1.34 & 2.70 & 1.98 \\
ULAS J0100+0620	& BRLT18 & 01:00:36 & +06:20:44 & 17.77 & 0.87 & 0.86 & 1.43 & - & - \\
ULAS J0105+1429	& BRLT20 & 01:05:32 & +14:29:32 & 18.01 & 1.25 & 0.54 & 1.18 & 2.59 & 2.33 \\
ULAS J0111-0105	& BRLT21 & 01:11:52 & -01:05:34 & 17.34 & 1.30 & 0.81 & 1.41 & 3.00 & 1.96 \\
ULAS J0112+1536	& BRLT22 & 01:12:50 & +15:36:58 & 18.00 & 1.01 & 0.59 & 1.14 & - & - \\
ULAS J0116+1443	& BRLT24 & 01:16:45 & +14:43:35 & 17.96 & 1.35 & 0.95 & 1.66 & - & - \\
SDSS J0127+1354$^{b}$	& BRLT26 & 01:27:44 & +13:54:21 & 16.77 & 1.19 & 0.86 & 1.59 & 2.85 & 2.57 \\
ULAS J0128-0041	& BRLT27 & 01:28:14 & -00:41:54 & 17.59 & 0.87 & 0.69 & 1.10 & 2.89 & 3.65 \\
ULAS J0132+0552	& BRLT30 & 01:32:44 & +05:52:32 & 16.41 & 1.35 & 0.93 & 1.66 & 2.89 & 2.03 \\
ULAS J0136+0717	& BRLT31 & 01:36:20 & +07:17:38 & 18.01 & 1.45 & 0.91 & 1.53 & - & - \\
ULAS J0138-0104	& BRLT32 & 01:38:08 & -01:04:17 & 18.01 & 1.31 & 0.67 & - & 2.81 & 1.59 \\
ULAS J0141+1318   &BRLT33 & 01:41:03 & +13:18:33 & 17.95 & 1.51 & 0.85 & 1.37 & 2.59 & 1.87 \\
ULAS J0148+1400	& BRLT35 & 01:48:12 & +14:00:28 & 17.97 & 1.12 & 0.81 & 1.42 & 3.52 & -0.16 \\
ULAS J0149+1441	& BRLT37 & 01:49:27 & +14:41:08 & 18.04 & 1.27 & 0.94 & 1.72 & 2.55 & 2.90 \\
ULAS J0150+1359$^{c}$ & - &01:50:24 & +13:59:24 & 17.73 & 1.08 & -0.38 & -0.12 & - & - \\
SDSS J0151+1244$^{a}$	& BRLT38 & 01:51:42 & +12:44:29 & 16.39 & 1.02 & 0.79 & 1.10 & - & - \\
ULAS J0151+1346	& BRLT39 & 01:51:44 & +13:46:46 & 17.66 & 1.24 & 0.82 & 1.57 & 2.62 & 2.89 \\
ULAS J0200+0658	& BRLT42 & 02:00:03 & +06:58:08 & 17.93 & 1.18 & 0.73 & 1.22 & - & - \\
SDSS J0203-0108$^{d}$	& BRLT44 & 02:03:33 & -01:08:12 & 17.69 & 1.30 & 0.81 & 1.42 & 2.77 & 3.56 \\
ULAS J0203-0102$^{e}$ &-  &02:03:36 & -01:02:31 & 18.05 & 1.09 & -0.29 & -0.11 & - & - \\
ULAS J0205+1421	& BRLT45 & 02:05:30 & +14:21:14 & 17.99 & 1.15 & 0.73 & 1.06 & 2.71 & 4.47 \\
ULAS J0206+0549	& BRLT46 & 02:06:04 & +05:49:59 & 17.92 & 1.06 & 0.50 & 1.11 & 2.48 & 1.92 \\
SDSS J0207+0000$^{a}$ & -& 02:07:42 & +00:00:56 & 16.73 & 1.29 & -0.07 & 0.01 & - & - \\
ULAS J0247-0107	& BRLT48 & 02:47:03 & -01:07:01 & 17.77 & 1.43 & 0.94 & 1.77 & - & - \\
ULAS J0255+0616$^{h}$	& BRLT50 & 02:55:45 & +06:16:56 & 17.99 & 1.16 & -0.68 & - & - & - \\
ULAS J0259+0549	& BRLT51 & 02:59:41 & +05:49:35 & 18.02 & 1.25 & 0.83 & 1.54 & - & - \\
ULAS J0314+0453	& BRLT52 & 03:14:52 & +04:53:46 & 17.30 & 1.29 & 0.91 & 1.71 & 3.06 & 2.00 \\
ULAS J0320+0617	& BRLT56 & 03:20:00 & +06:17:41 & 17.79 & 1.46 & 0.75 & 1.39 & - & - \\
ULAS J0320+0618	& BRLT57 & 03:20:42 & +06:18:37 & 18.06 & 1.21 & 0.58 & 1.15 & 2.43 & 1.93 \\
ULAS J0321+0545	& BRLT58 & 03:21:43 & +05:45:24 & 17.33 & 1.22 & 0.73 & 1.37 & 2.74 & 2.34 \\
ULAS J0323+0613	& BRLT60 & 03:23:54 & +06:13:52 & 17.64 & 1.37 & 0.65 & 1.33 & - & - \\
SDSS J0325+0425$^{f}$ & - & 03:25:53 & +04:25:40 & 16.02 & 1.10 & -0.22 & -0.43 & - & - \\
ULAS J0330+0556	& BRLT62 & 03:30:06 & +05:56:53 & 17.95 & 1.56 & 1.10 & 2.00 & - & - \\
ULAS J0330+0426	& BRLT64 & 03:30:37 & +04:26:58 & 17.29 & 1.33 & 0.85 & 1.54 & 2.69 & 2.16 \\
ULAS J0341+0423	& BRLT66 & 03:41:50 & +04:23:25 & 16.85 & 1.44 & 0.90 & 1.65 & 2.93 & 2.12 \\
ULAS J2157+0056	& BRLT305 & 21:57:00 & +00:56:15 & 17.85 & 1.41 & 0.98 & 1.75 & 2.77 & 1.66 \\
ULAS J2159+0033	& BRLT306 & 21:59:20 & +00:33:10 & 17.73 & 1.36 & 0.74 & 1.37 & 2.76 & 2.51 \\
ULAS J2209-0053   &BRLT307 & 22:09:17 & -00:53:00 & 18.01 & 1.34 & 0.74 & 1.37 & 2.48 & 2.31 \\
ULAS J2229+0102	 &BRLT311 & 22:29:58 & +01:02:17 & 17.88 & 1.22 & 0.39 & 0.67 & - & - \\
ULAS J2233+0022	 &BRLT312 & 22:33:48 & +00:22:14 & 18.07 & 1.05 & 0.71 & 1.43 & 3.45 & 0.42 \\
ULAS J2236+0111	 &BRLT313 & 22:36:37 & +01:11:32 & 17.11 & 1.34 & 0.87 & 1.64 & 3.04 & 2.03 \\
ULAS J2237+0716	 &BRLT314 & 22:37:57 & +07:16:57 & 17.49 & 1.38 & 1.04 & 1.84 & 2.97 & 2.73 \\
ULAS J2240+0008	 &BRLT315 & 22:40:52 & +00:08:22 & 17.82 & 1.46 & 0.70 & 1.24 & 2.44 & 2.03 \\
ULAS J2249+0715	 &BRLT316 & 22:49:23 & +07:15:28 & 18.09 & 1.55 & 0.55 & 1.23 & - & - \\
ULAS J2250+0808	 &BRLT317 & 22:50:16 & +08:08:22 & 15.50 & 1.17 & 0.46 & 0.99 & 2.74 & 2.11 \\
ULAS J2251-0007	 &BRLT318 & 22:51:15 & -00:07:24 & 17.95 & 1.26 & 0.60 & 1.46 & 3.20 & 0.97 \\
ULAS J2256+0724	 &BRLT320 & 22:56:31 & +07:24:39 & 17.94 & 1.48 & 0.68 & 1.21 & 2.71 & 2.45 \\
ULAS J2302+0700	 &BRLT321 & 23:02:03 & +07:00:39 & 17.62 & 1.33 & 0.25 & 0.11 & - & - \\
ULAS J2303+0058	 &BRLT322 & 23:03:59 & +00:58:07 & 17.82 & 1.21 & 0.83 & 1.67 & 2.86 & 2.59 \\
ULAS J2304+1301	 &BRLT323 & 23:04:25 & +13:01:11 & 16.69 & 1.31 & 0.77 & 1.49 & 2.78 & 2.06 \\
ULAS J2304+0804	 &BRLT325 & 23:04:34 & +08:04:01 & 17.89 & 1.23 & 0.41 & 0.67 & - & - \\
\hline
\end{tabular}
\label{photom}
\end{table*}

\begin{table*}
\raggedright{\bf Table~\ref{photom} continued.} Photometry of sub-sample. \\
\centering
\begin{tabular}{|l|l|c|c|c|c|c|c|c|c|}
\hline
Name &ID & RA & DEC & J$^1$ & Y-J$^1$ & J-H$^1$ & J-K$^1$ & z-J$^{1,2}$ & i-z$^2$ \\
\hline
ULAS J2306+1302$^{c}$ &  -& 23:06:01 & +13:02:25 & 17.57 & 1.39 & -0.43 & -0.46 & - & - \\
ULAS J2312+0006	 &BRLT328 & 23:12:37 & +00:06:02 & 17.65 & 1.30 & 0.60 & 1.25 & 2.62 & 2.07 \\
ULAS J2316+0100	 &   BRLT330 & 23:16:46 & +01:00:13 & 17.95 & 1.15 & 0.69 & 1.25 & 3.13 & 1.94 \\
ULAS J2320+1448$^{g}$ & - & 23:20:35 & +14:48:29 & 16.79 & 1.35 & -0.35 & -0.61 & - & - \\
ULAS J2321-0045	 &   BRLT331 & 23:21:23 & -00:45:57 & 18.00 & 1.40 & 0.40 & 0.88 & 2.52 & 2.44 \\
ULAS J2321+1354$^{g}$ & -& 23:21:23 & +13:54:54 & 16.72 & 1.20 & -0.43 & -0.44 & - & - \\
ULAS J2323+0005	 &   BRLT332 & 23:23:00 & +00:05:42 & 18.01 & 1.15 & 0.74 & 1.15 & 2.53 & 2.17 \\
ULAS J2323+0719	 &   BRLT333 & 23:23:15 & +07:19:31 & 17.30 & 1.20 & 0.75 & 1.10 & 3.05 & 3.38 \\
ULAS J2327+1517	 &   BRLT334 & 23:27:16 & +15:17:30 & 16.20 & 1.34 & 0.85 & 1.52 & 2.97 & 2.13 \\
ULAS J2327+0102	 &   BRLT335 & 23:27:32 & +01:02:53 & 18.07 & 1.19 & 0.83 & 1.46 & - & - \\
ULAS J2328+1345$^{c}$ & -& 23:28:02 & +13:45:44 & 17.75 & 1.26 & -0.42 & -0.54 & - & - \\
ULAS J2330+1403    & BRLT338 & 23:30:02 & +14:03:30 & 17.37 & 1.23 & 0.57 & 1.26 & - & - \\
ULAS J2347-0110     &BRLT343 & 23:47:17 & -01:10:09 & 17.57 & 1.25 & 0.85 & 1.67 & 2.70 & 2.35 \\
ULAS J2356+0754     &BRLT344 & 23:56:18 & +07:54:20 & 18.09 & 1.51 & 1.10 & 1.87 & - & - \\
\hline
\multicolumn{10}{l}{$^1$: Colours and photometry from UKIDSS \citep{hewett06}, $^2$: Colours and photometry from SDSS \citep{fukugita96}.}\\
\multicolumn{10}{l}{Original discovery paper: $^{a}$\citet{geballe02}, $^{b}$\citet{hawley02}, $^{c}$\citet{burningham10a}, $^{d}$\citet{knapp04}, }\\
\multicolumn{10}{l}{$^{e}$\citet{lodieu07a}, $^{f}$\citet{chiu06}, $^{g}$\citet{scholz10}, $^{h}$Burningham et al., in prep.}\\
\hline
\end{tabular}
\end{table*}

\section{Observations \& Data reduction}
\label{obs}
Spectroscopic observations of candidates from our sub-sample, were obtained with X-shooter on the Very Large Telescope during Nov 27-31 2010, Feb 22-26, June 5-9 2011 and Sep 18-22 2011, under the ESO programs 086.C-0450(A/B) and 087.C-0639(A/B). We used the echelle slit mode, which covers the wavelength range 300-2500nm. This is split into three separate arms, the UVB (300-550nm), VIS (550-1000nm) and NIR (1000-2500nm). Using slit widths of 1.0 arcsec for the UVB arm and 0.9 arcsec for the VIS and NIR arms, we took four individual integrations in an ABBA pattern (See Table.~\ref{obsdetails} for integration times). We note that we do not detect any significant flux in the UVB wavelength range for members of our sample and as such we do not show the details for this arm. We took telluric standard stars after every second target, which were paired together so they were in roughly the same airmass. Sky flats and arc frames were also taken at the beginning of every night.

The data were reduced using the ESO X-shooter pipeline (version 1.3.7). The pipeline removes non-linear pixels, subtracts the bias (in the VIS arm) or dark frames (in the NIR arm) and divides the raw frames by flat fields. Images are pair-wise subtracted to remove sky background. The pipeline then extracts and merges the different orders in each arm, rectifying them using a multi-pinhole arc lamp (taken during the day-time calibration) and correcting for the flexure of the instrument using single-pinhole arc lamps (taken at night, one for each object observed). Telluric stars are reduced in the same way, except that sky subtraction is done by fitting the background (as tellurics are not observed in nodding mode). The spectra were telluric corrected and flux calibrated using IDL routines, following a standard procedure: first the telluric spectrum is cleared of HI absorption lines (by interpolating over them) and scaled to match the measured magnitudes; then is divided by a blackbody curve for the appropriate temperature, to obtain the instrument$+$atmosphere response curve; finally the target spectra is multiplied by the response curve obtained to flux calibrate it. The arms (VIS and NIR) were then merged by matching the flux level in the overlapping regions between them. The flux calibration was checked by determining the target's synthetic MKO $YJHK$ magnitudes, that were compared to those obtained in the ULAS. Finally, each spectrum was visually inspected to check for possible problems during the extraction or merging stage. The spectra were then binned (in the $\lambda$ direction) by 40 times to produce an average S/N$=30$ for resolution R$=$ 880 and 510 in the VIS and NIR arms, respectively.

\begin{table*}
\caption{Optical and near-infrared spectroscopic observations of sub-sample.} 
\centering
\begin{tabular}{|l|c|c|c|c|c|}
\hline
Name & UT date & VIS $t_{int}$(s) & VIS S/N$^{*}$ & NIR $t_{int}$(s) & NIR S/N$^{*}$ \\
\hline
BRLT1	&	19-Sep-2011	&	1600	&	4.54	&	1960	&	29.7	\\
BRLT2	&	20-Sep-2011	&	2000	&	4.72	&	2360	&	18.8	\\
BRLT3	&	28-Nov-2010	&	1200	&	5.21	&	1560	&	32.8	\\
BRLT6	&	18-Sep-2011	&	2000	&	5.32	&	2360	&	25.2	\\
BRLT7	&	20-Sep-2011	&	1600	&	7.61	&	1960	&	20.9	\\
BRLT8	&	21-Sep-2011	&	1600	&	5.18	&	1960	&	31	\\
BRLT9	&	19-Sep-2011	&	1600	&	8.01	&	1960	&	30.7	\\
BRLT10	&	29-Nov-2010	&	800	&	2.39	&	1160	&	26.5	\\
BRLT12	&	21-Sep-2011	&	2000	&	4.3	&	2360	&	18.7	\\
BRLT14	&	30-Nov-2010	&	1600	&	7.56	&	1960	&	29.8	\\
BRLT15	&	20-Sep-2011	&	1600	&	2.4	&	1960	&	21.6	\\
BRLT16	&	28-Nov-2010	&	1400	&	3.66	&	1760	&	22.8	\\
BRLT18	&	30-Nov-2010	&	1600	&	5.79	&	1960	&	29	\\
BRLT20	&	21-Sep-2011	&	2000	&	5.12	&	2360	&	20.7	\\
BRLT21	&	19-Sep-2011	&	1200	&	3.73	&	800	&	29.5	\\
BRLT22	&	21-Sep-2011	&	1600	&	5.84	&	1960	&	17.7	\\
BRLT24	&	21-Sep-2011	&	1600	&	4.17	&	1960	&	23.9	\\
BRLT26	&	27-Nov-2010	&	600	&	4.14	&	600	&	24.2	\\
BRLT27	&	29-Nov-2010	&	1600	&	2.22	&	1960	&	27.6	\\
BRLT30	&	28-Nov-2010	&	800	&	6.39	&	1160	&	53.3	\\
BRLT31	&	18-Sep-2011	&	2000	&	6.18	&	2360	&	27.6	\\
BRLT32	&	20-Sep-2011	&	2000	&	5.76	&	2360	&	25	\\
BRLT33	&	19-Sep-2011	&	1600	&	2.93	&	1960	&	20.9	\\
BRLT35	&	21-Sep-2011	&	1600	&	5.86	&	1960	&	21.6	\\
BRLT37	&	18-Sep-2011	&	2000	&	4.84	&	2360	&	26	\\
BRLT38	&	27-Nov-2010	&	240	&	4.14	&	600	&	31.9	\\
BRLT39	&	18-Sep-2011	&	1600	&	7.77	&	1960	&	29.8	\\
BRLT42	&	20-Sep-2011	&	1600	&	5.75	&	1960	&	23.6	\\
BRLT44	&	30-Nov-2010	&	1600	&	4.3	&	1960	&	28.6	\\
BRLT45	&	21-Sep-2011	&	1600	&	3.76	&	1960	&	16.7	\\
BRLT46	&	30-Nov-2010	&	1600	&	3.84	&	1960	&	21.6	\\
BRLT48	&	28-Nov-2010	&	1400	&	2.93	&	1760	&	27.6	\\
BRLT50	&	30-Nov-2010	&	2400	& $\sim$1 	&	2760	&	7.26	\\
BRLT51	&	19-Sep-2011	&	2000	&	4.59	&	2360	&	22.9	\\
BRLT52	&	27-Nov-2010	&	1200	&	6.55	&	1200	&	29.1	\\
BRLT56	&	29-Nov-2010	&	1600	&	3.01	&	1960	&	22.8	\\
BRLT57	&	30-Nov-2010	&	2400	&	3.93	&	2760	&	26.6	\\
BRLT58	&	27-Nov-2010	&	1200	&	5.09	&	1200	&	27.1	\\
BRLT60	&	25-Feb-2011	&	1400	&	4.11	&	1760	&	27.9	\\
BRLT62	&	29-Nov-2010	&	1600	&	2.64	&	1960	&	26.8	\\
BRLT64	&	27-Nov-2010	&	1200	&	5.26	&	1200	&	29.5	\\
BRLT66	&	27-Nov-2010	&	800	&	4.82	&	800	&	27.5	\\
BRLT305	&	8-Jun-2011	&	1600	&	1.94	&	1960	&	19.3	\\
BRLT306	&	7-Jun-2011	&	1600	&	1.71	&	1960	&	25.7	\\
BRLT307	&	18-Sep-2011	&	2000	&	7.84	&	2360	&	27.2	\\
BRLT311	&	19-Sep-2011	&	1600	&	2.93	&	1960	&	16.2	\\
BRLT312	&	20-Sep-2011	&	2000	&	4.38	&	2360	&	22.9	\\
BRLT313	&	25-Feb-2011	&	1600	&	6.34	&	1960	&	42	\\
BRLT314	&	7-Jun-2011	&	1200	&	1.71	&	1560	&	28.2	\\
BRLT315	&	19-Sep-2011	&	1600	&	4.96	&	1960	&	25.3	\\
BRLT316	&	21-Sep-2011	&	2000	&	4.41	&	2360	&	20.3	\\
BRLT317	&	28-Nov-2010	&	800	&	13.7	&	1160	&	77.1	\\
BRLT318	&	18-Sep-2011	&	1600	&	4.91	&	1960	&	21.7	\\
BRLT320	&	20-Sep-2011	&	1600	&	4.49	&	1960	&	19.7	\\
BRLT321	&	8-Jun-2011	&	1600	&	1.26	&	1960	&	16	\\
BRLT322	&	18-Sep-2011	&	1600	&	5.48	&	1960	&	27.7	\\
BRLT323	&	29-Nov-2010	&	800	&	3.83	&	1160	&	37.8	\\
BRLT325	&	19-Sep-2011	&	1600	&	2.39	&	1960	&	18	\\
BRLT328	&	29-Nov-2010	&	1600	&	4.23	&	1960	&	23.4	\\
\hline
\end{tabular}
\label{obsdetails}
\end{table*}

\begin{table*}
\raggedright{\bf Table~\ref{obsdetails} continued.} Optical and near-infrared spectroscopic observations of sub-sample.\\
\centering
\begin{tabular}{|l|c|c|c|c|c|}
\hline
BRLT330	&	20-Sep-2011	&	1600	&	5.2	&	1960	&	20.6	\\
BRLT331	&	19-Sep-2011	&	2000	&	5.45	&	2360	&	21.9	\\
BRLT332	&	18-Sep-2011	&	2000	&	7.87	&	2360	&	27.7	\\
BRLT333	&	28-Nov-2010	&	800	&	1.7	&	1160	&	21.5	\\
BRLT334	&	27-Oct-2010	&	240	&	4.06	&	600	&	33.9	\\
BRLT335	&	21-Sep-2011	&	2000	&	3.75	&	2360	&	24.5	\\
BRLT338	&	27-Nov-2010	&	800	&	3.44	&	800	&	18.2	\\
BRLT343	&	28-Nov-2010	&	1200	&	3.67	&	1560	&	28	\\
BRLT344	&	18-Sep-2011	&	2000	&	4.17	&	2360	&	26.8	\\
\hline
\multicolumn{6}{l}{$^{*}$S/N is per spectral pixel, after binning.}\\
\hline
\end{tabular}
\end{table*}

\section{Spectroscopy}
\label{spec}
\subsection{Spectral Types}
\label{spec_types}
\begin{figure}
\includegraphics[width=90mm,angle=0]{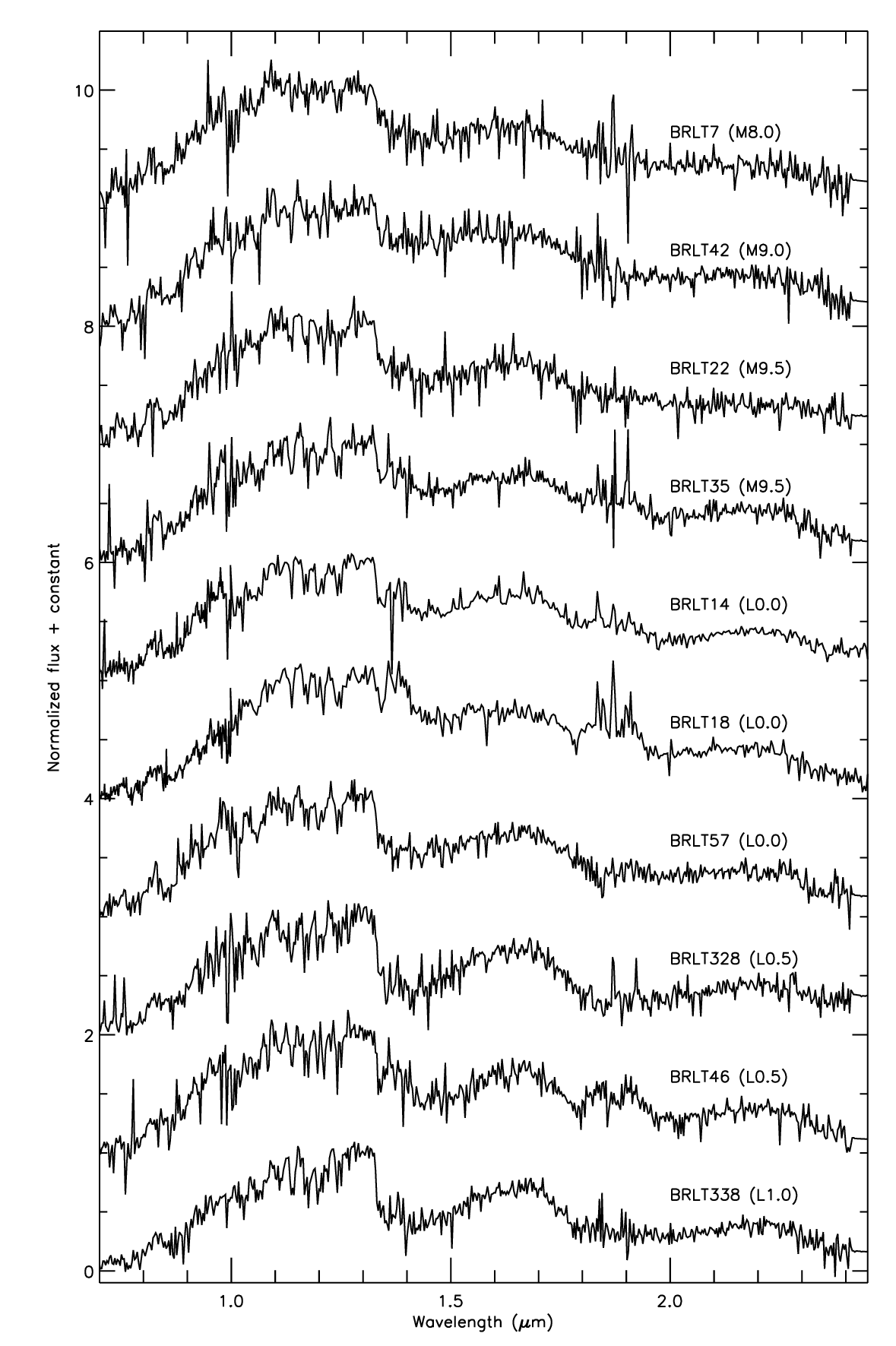}
\caption{X-shooter spectra of L and T dwarfs from birth rate sample, in ascending spectral type. }
\label{spectra1}
\end{figure}

\begin{figure}
\includegraphics[width=90mm,angle=0]{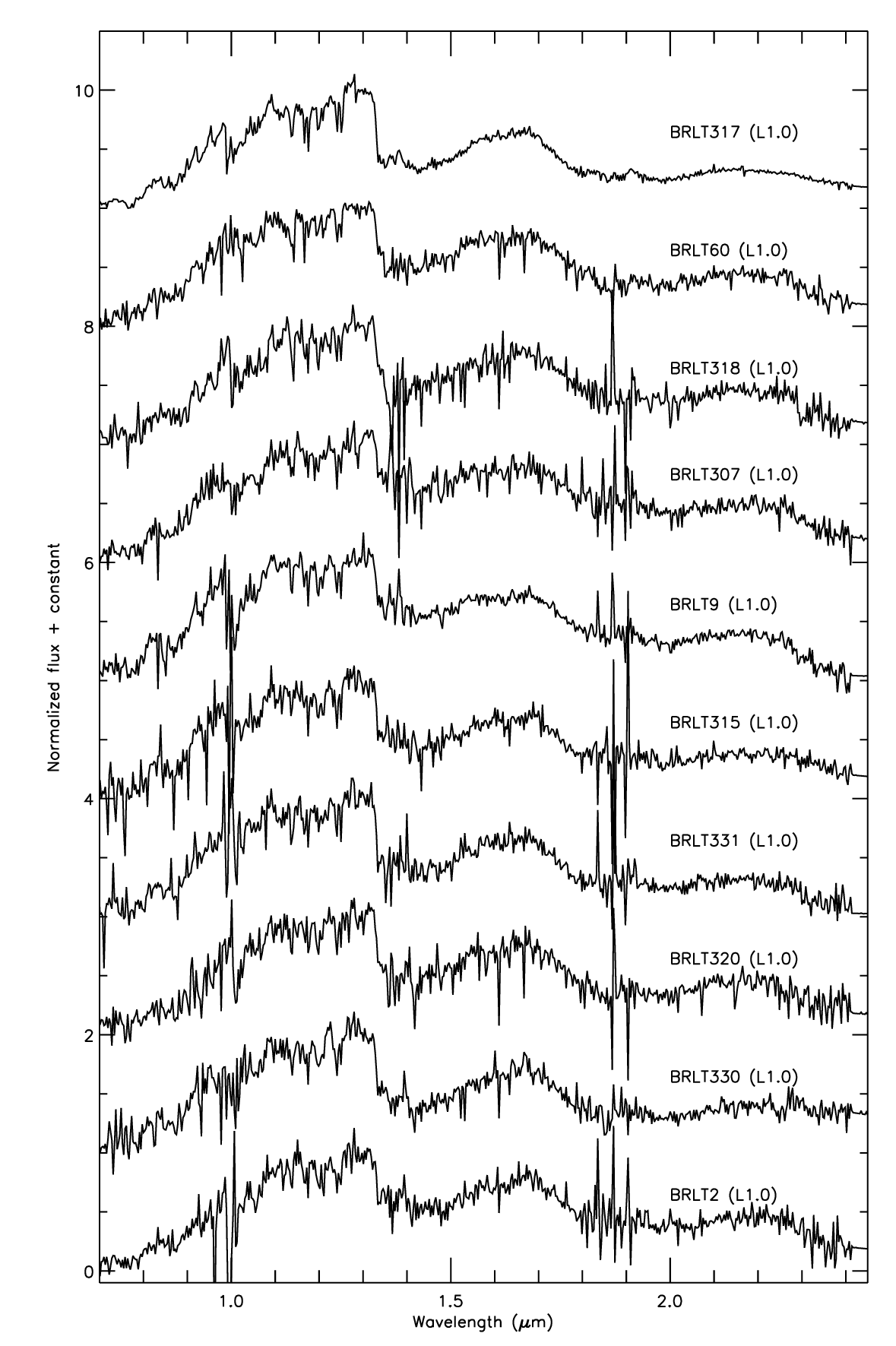}
\caption{X-shooter spectra of L and T dwarfs from birth rate sample, in ascending spectral type.}
\label{spectra2}
\end{figure}

\begin{figure}
\includegraphics[width=90mm,angle=0]{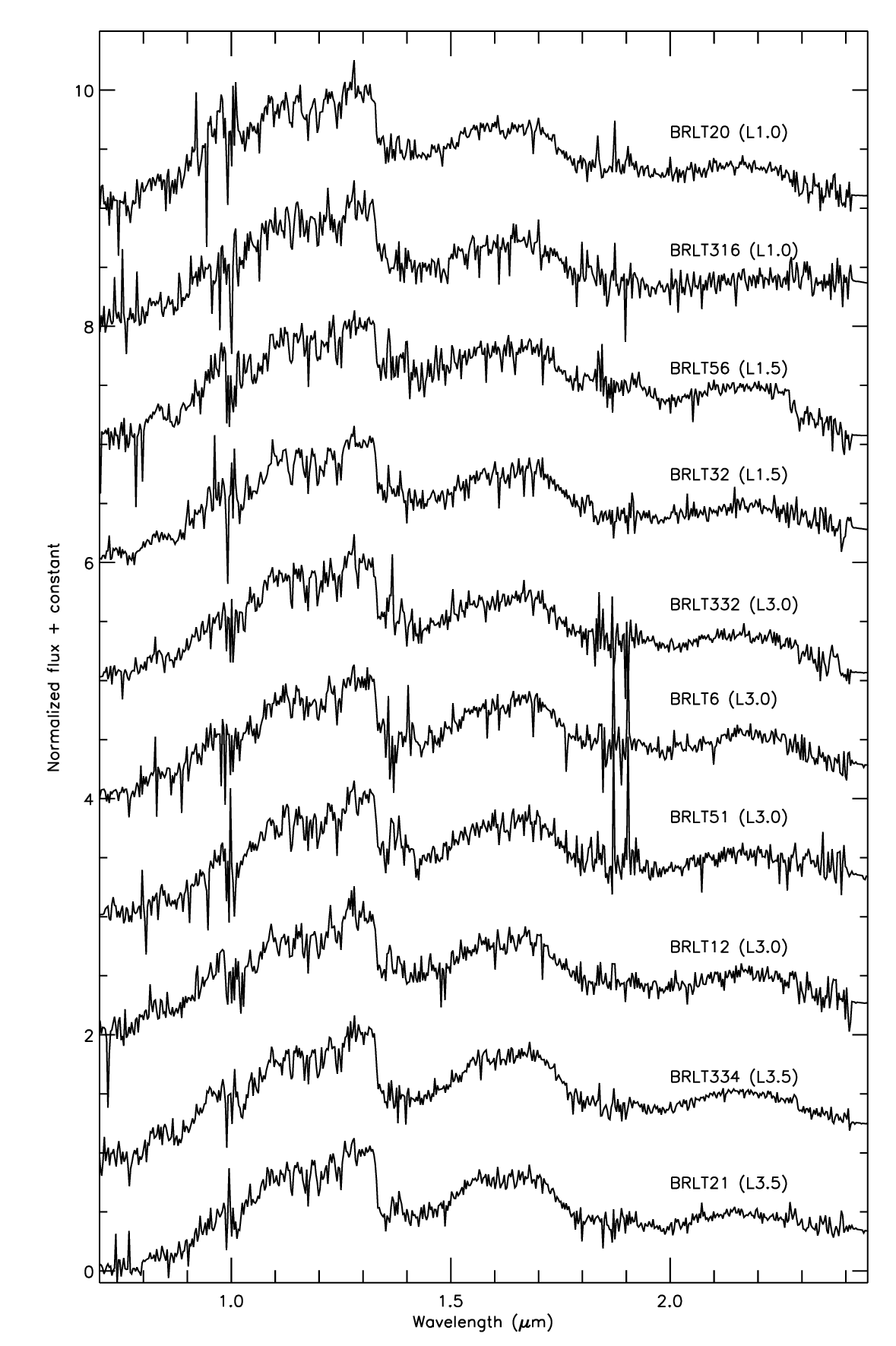}
\caption{X-shooter spectra of L and T dwarfs from birth rate sample, in ascending spectral type.}
\label{spectra3}
\end{figure}

\begin{figure}
\includegraphics[width=90mm,angle=0]{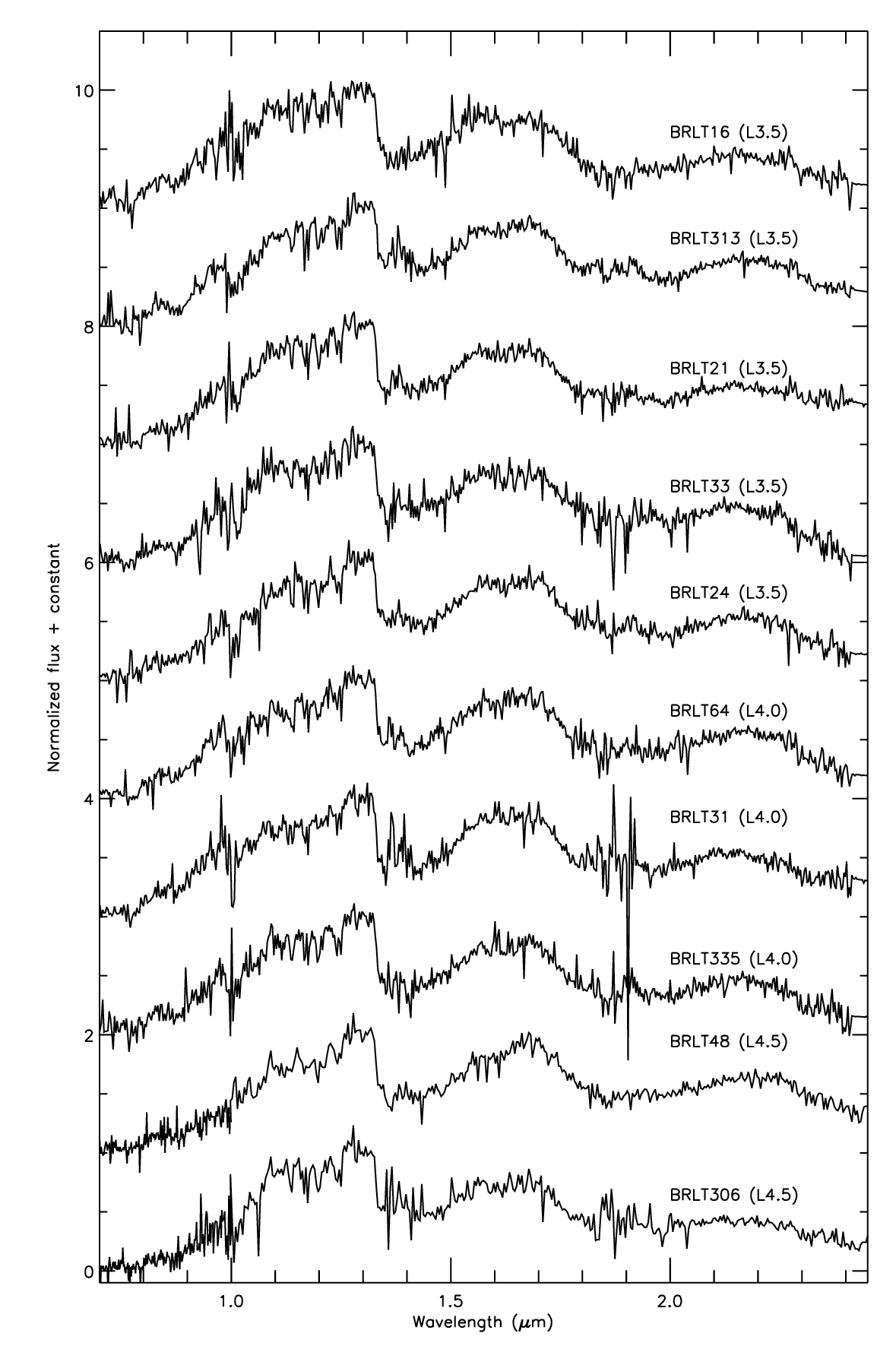}
\caption{X-shooter spectra of L and T dwarfs from birth rate sample, in ascending spectral type.}
\label{spectra4}
\end{figure}

\begin{figure}
\includegraphics[width=90mm,angle=0]{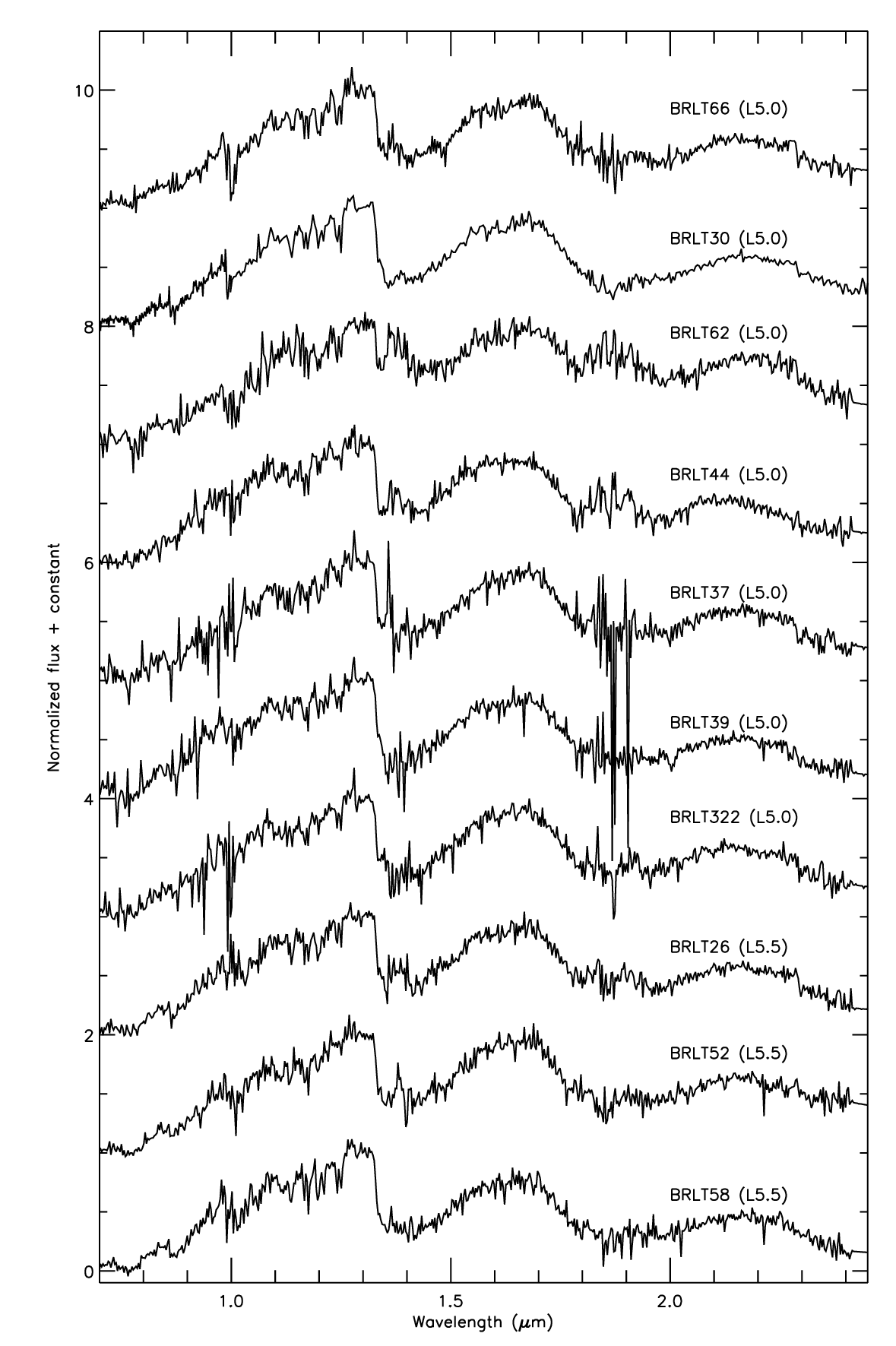}
\caption{X-shooter spectra of L and T dwarfs from birth rate sample, in ascending spectral type.}
\label{spectra5}
\end{figure}

\begin{figure}
\includegraphics[width=90mm,angle=0]{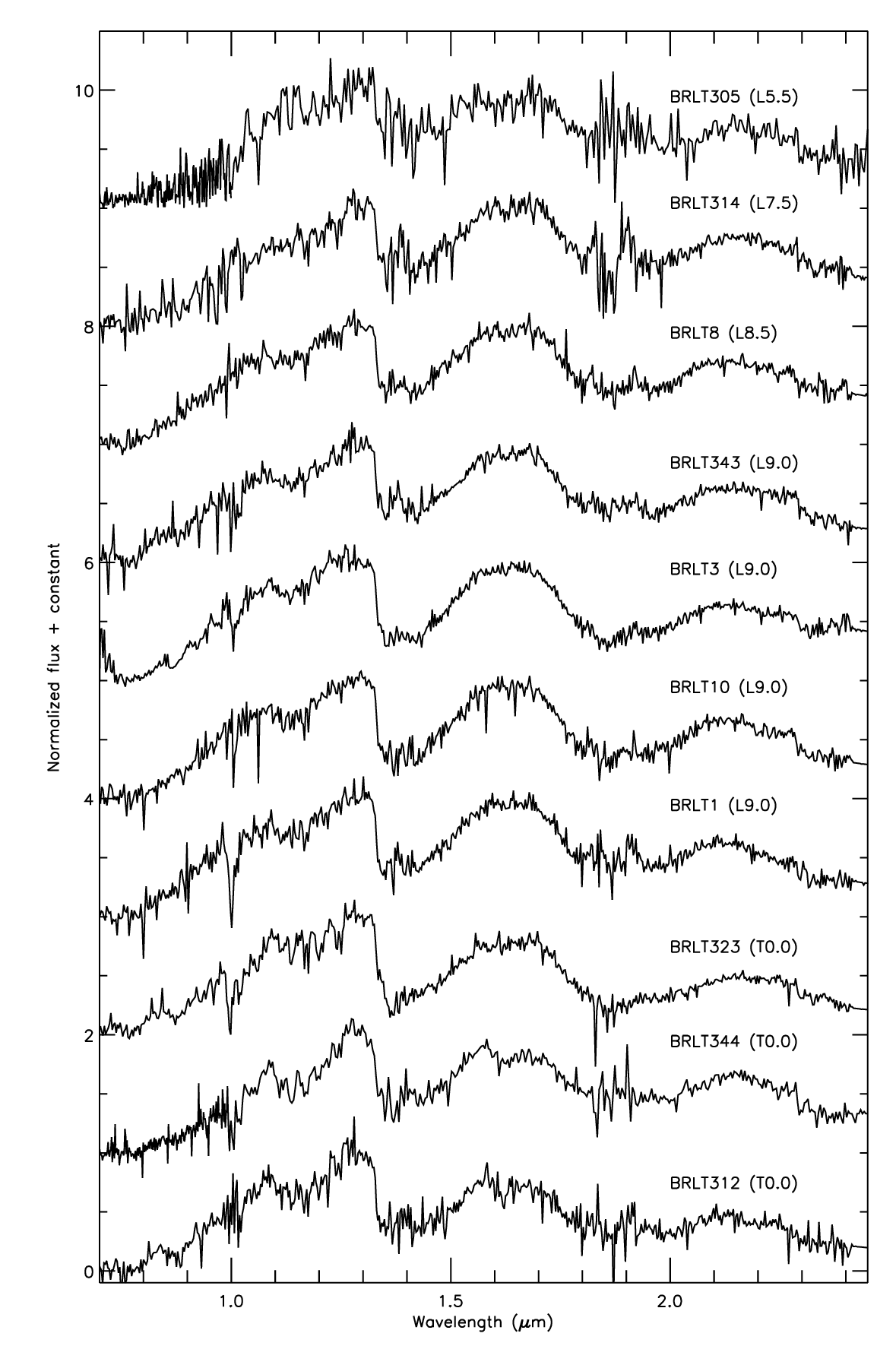}
\caption{X-shooter spectra of L and T dwarfs from birth rate sample, in ascending spectral type.}
\label{spectra6}
\end{figure}

\begin{figure}
\includegraphics[width=90mm,angle=0]{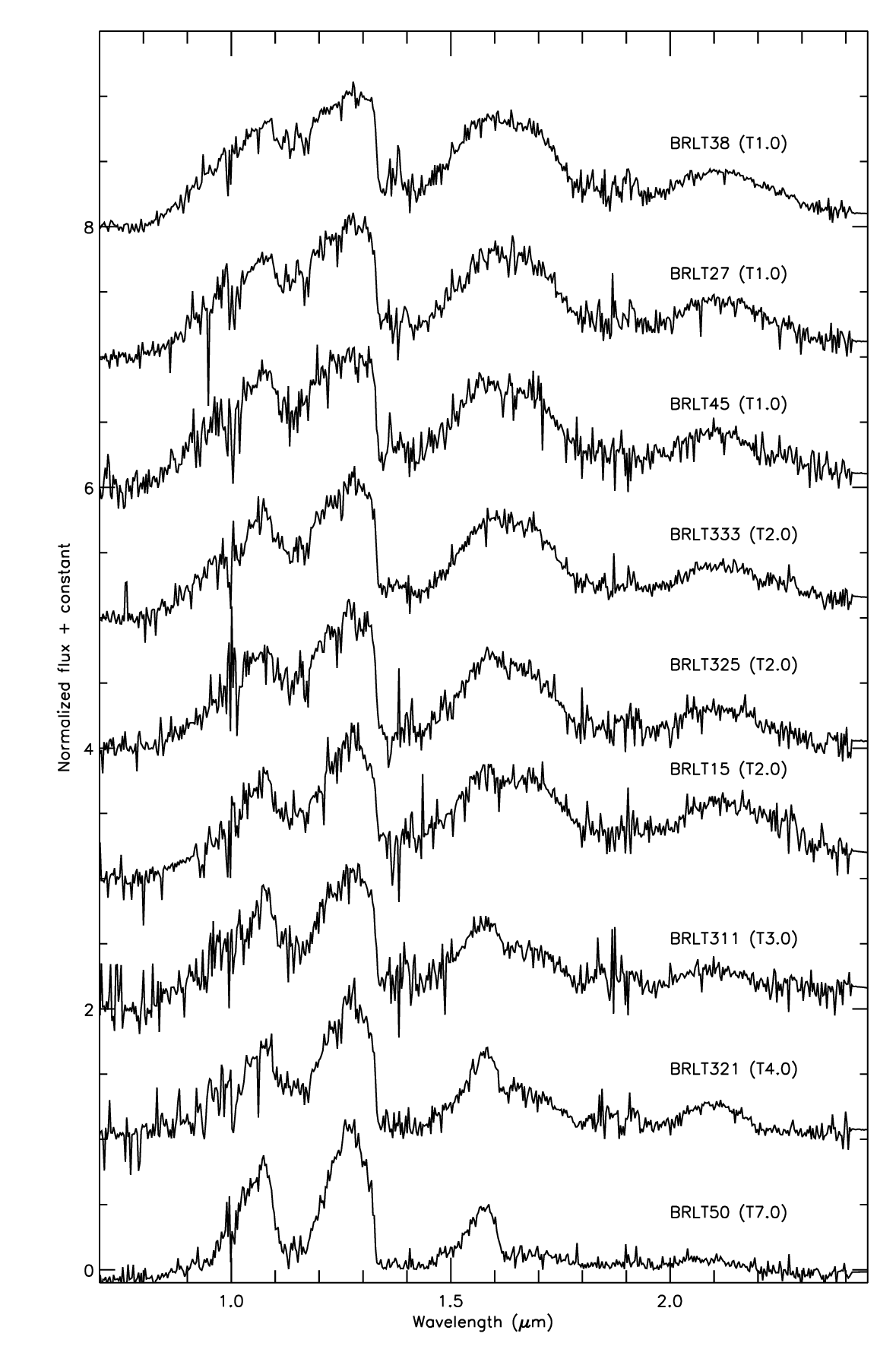}
\caption{X-shooter spectra of L and T dwarfs from birth rate sample, in ascending spectral type.}
\label{spectra7}
\end{figure}

Spectral types were determined via spectral fitting with template objects. Template spectra were obtained from the SpeX Prism Spectral Library \footnote{http://pono.ucsd.edu/~adam/browndwarfs/spexprism/}, which are shown in Table.~\ref{spt}. Our observed spectra were then re-sampled to the same resolution as the SpeX data (see Table.~\ref{spt}) and spectral types were determined from the best chi-squared fit to the templates using the spectral typing scheme from \citet{kp99} and \citet{burgy06a}, for L and T dwarfs, respectively. The three best fits were also visually inspected to make sure that they were good fits to the real spectra. The spectral types, as well as their associated errors are shown in Table.~\ref{specinfo} and the spectra can be seen in ascending spectral type in Fig.~\ref{spectra1}~-~\ref{spectra7}. 

\begin{table*}
\caption{Spectral templates from the SpeX Prism Spectral Library used for spectral typing.}
\begin{tabular}{|l|l|l|c|c|c|}
\hline
Object Name & 2MASS Designation & Optical SpT & NIR SpT & Resolution & Reference \\
\hline
VB 8                      & J16553529-0823401 & M7 V & M7 & 120 & 1 \\
VB 10                     & J19165762+0509021 & M8 V & M8 & 120 & 2 \\
LHS 2924                  & J14284323+3310391 & M9 V & M9 & 120 &3 \\
2MASSW J2130446-084520    & J21304464-0845205 & L1.5 & L1 & 120 &4 \\
Kelu-1                    & J13054019-2541059 & L2   & L2 & 120 &5 \\
2MASSW J1506544+132106    & J15065441+1321060 & L3   & L3 & 120 &6 \\
2MASS J21580457-1550098   & J21580457-1550098 & L4   & L4 & 120 & 4 \\
SDSS J083506.16+195304.4  & J08350622+1953050 & -    & L5 & 120 & 7 \\
2MASSI J1010148-040649    & J10101480-0406499 & L6   & L6 & 120 &8 \\
2MASSI J0103320+193536    & J01033203+1935361 & L6   & L7 & 120 &9 \\
2MASSW J1632291+190441    & J16322911+1904407 & L8   & L8 &  75 & 5 \\
DENIS-P J0255-4700        & J02550357-4700509 & L8   & L9 & 120 & 10 \\
SDSS J120747.17+024424.8  & J12074717+0244249 & L8   & T0 & 120 &11 \\
SDSS J015141.69+124429.6  & J01514155+1244300 & -    & T1 & 120 & 12 \\
SDSSp J125453.90-012247.4 & J12545393-0122474 & T2   & T2 & 120 & 12\\
2MASS J12095613-1004008   & J12095613-1004008 & T3.5 & T3 & 120 & 12\\
2MASSI J2254188+312349    & J22541892+3123498 & -    & T4 & 120 & 12\\
2MASS J15031961+2525196   & J15031961+2525196 & T6   & T5 & 120 & 12\\
SDSSp J162414.37+002915.6 & J16241436+0029158 & -    & T6 & 120 & 13\\
2MASSI J0727182+171001    & J07271824+1710012 & T8   & T7 & 120 & 13\\
2MASSI J0415195-093506    & J04151954-0935066 & T8   & T8 & 120 & 12\\
\hline
\multicolumn{6}{l}{1:~\citet{burgy08}, 2:~\citet{burgy04}, 3:~\citet{burgy06}, 4:~\citet{kp10}, }\\
\multicolumn{6}{l}{5:~\citet{burgy07}, 6:~\citet{burgy07b}, 7:~\citet{chiu06}, 8:~\citet{reid06}, }\\
\multicolumn{6}{l}{9:~\citet{cruz04}, 10:~\citet{burgy10}, 11:~\citet{looper07}, }\\
\multicolumn{6}{l}{12:~\citet{burgy04}, 13:~\citet{burgy06a} }\\
\hline
\end{tabular}
\label{spt}
\end{table*}

\begin{table*}
\small
\caption{Spectral details of observed targets.}
\centering
\begin{tabular}{|l|c|c|c|c|c|c|c|c|c|c|}
\hline
ID & Spectral type & H$_2$O-J & H$_2$O-H & H$_2$O-K & CH$_4$-J & CH$_4$-H & CH$_4$-K & K/J & H-dip & Notes \\
\hline
BRLT1 &  L9.0 $\pm$ 0.5 & 0.72 & 0.68 & 0.86 & 0.75 & 1.00 & 0.76 & 0.60 & 0.49 & \\
BRLT2 &  L1.0 $\pm$ 1.0 & 0.94 & 0.76 & 1.13 & 0.80 & 1.09 & 1.06 & 0.40 & 0.46 & \\
BRLT3 &  L9.0 $\pm$ 1.0 & 0.75 & 0.69 & 0.87 & 0.73 & 1.04 & 0.90 & 0.58 & 0.50 & \\
BRLT6 &  L3.0 $\pm$ 1.0 & 0.85 & 0.82 & 1.04 & 0.83 & 1.13 & 1.11 & 0.45 & 0.49 & \\
BRLT7 &  M8.0 $\pm$ 1.0 & 0.95 & 0.87 & 1.15 & 0.86 & 0.98 & 0.91 & 0.36 & 0.50 & \\
BRLT8 &  L8.5 $\pm$ 0.5 & 0.71 & 0.75 & 0.87 & 0.77 & 1.03 & 0.89 & 0.67 & 0.49 & \\
BRLT9 &  L1.0 $\pm$ 1.0 & 0.95 & 0.84 & 0.98 & 0.86 & 1.03 & 0.95 & 0.36 & 0.48 & \\
BRLT10 &  L9.0 $\pm$ 0.5 & 0.67 & 0.72 & 0.79 & 0.76 & 1.05 & 0.85 & 0.64 & 0.52 & \\
BRLT12 &  L3.0 $\pm$ 1.0 & 0.75 & 0.76 & 0.95 & 0.74 & 1.05 & 0.99 & 0.45 & 0.47 & \\
BRLT14 &  L0.0 $\pm$ 0.5 & 0.97 & 0.84 & 1.13 & 0.86 & 1.05 & 1.05 & 0.37 & 0.48 & \\
BRLT15 &  T2.0 $\pm$ 2.0 & 0.49 & 0.63 & 0.81 & 0.57 & 0.92 & 0.75 & 0.52 & 0.47 & SC \\
BRLT16 &  L3.5 $\pm$ 0.5 & 0.90 & 0.75 & 0.96 & 0.83 & 0.93 & 0.95 & 0.42 & 0.45 & WC \\
BRLT18 &  L0.0 $\pm$ 1.0 & 1.03 & 0.91 & 1.02 & 0.94 & 1.03 & 0.95 & 0.39 & 0.49 & \\
BRLT20 &  L1.0 $\pm$ 1.0 & 0.87 & 0.76 & 1.00 & 0.76 & 0.99 & 0.91 & 0.32 & 0.47 & \\
BRLT21 &  L3.5 $\pm$ 0.5 & 0.83 & 0.75 & 0.95 & 0.80 & 1.00 & 0.98 & 0.45 & 0.47 & \\
BRLT22 &  M9.5 $\pm$ 0.5 & 1.01 & 0.90 & 1.16 & 0.87 & 1.12 & 0.86 & 0.32 & 0.51 & \\
BRLT24 &  L3.5 $\pm$ 0.5 & 0.80 & 0.74 & 0.96 & 0.80 & 1.00 & 0.97 & 0.47 & 0.46 & \\
BRLT26 &  L5.5 $\pm$ 0.5 & 0.80 & 0.72 & 0.91 & 0.81 & 1.07 & 0.96 & 0.53 & 0.48 & \\
BRLT27 &  T1.0 $\pm$ 0.5 & 0.62 & 0.64 & 0.78 & 0.63 & 1.01 & 0.59 & 0.40 & 0.50 & \\
BRLT30 &  L5.0 $\pm$ 0.5 & 0.74 & 0.75 & 0.95 & 0.79 & 1.08 & 1.01 & 0.51 & 0.48 & \\
BRLT31 &  L4.0 $\pm$ 1.0 & 0.74 & 0.69 & 0.92 & 0.77 & 1.01 & 0.85 & 0.51 & 0.50 & \\
BRLT32 &  L1.5 $\pm$ 0.5 & 0.85 & 0.79 & 1.08 & 0.83 & 1.04 & 1.05 & 0.40 & 0.48 & \\
BRLT33 &  L3.5 $\pm$ 0.5 & 0.82 & 0.73 & 0.90 & 0.80 & 0.94 & 0.91 & 0.45 & 0.47 & WC \\
BRLT35 &  M9.5 $\pm$ 0.5 & 1.01 & 0.85 & 1.00 & 0.93 & 1.04 & 1.00 & 0.41 & 0.48 & \\
BRLT37 &  L5.0 $\pm$ 0.5 & 0.71 & 0.71 & 0.82 & 0.76 & 1.09 & 0.93 & 0.53 & 0.50 & \\
BRLT38 &  T1.0 $\pm$ 0.5 & 0.65 & 0.65 & 0.77 & 0.68 & 0.95 & 0.67 & 0.40 & 0.51 & \\
BRLT39 &  L5.0 $\pm$ 1.0 & 0.79 & 0.75 & 0.93 & 0.77 & 1.04 & 0.95 & 0.46 & 0.50 & \\
BRLT42 &  M9.0 $\pm$ 0.5 & 1.04 & 0.89 & 1.10 & 0.95 & 0.94 & 0.94 & 0.45 & 0.47 & WC \\
BRLT44 &  L5.0 $\pm$ 1.0 & 0.72 & 0.75 & 0.83 & 0.74 & 1.00 & 0.81 & 0.52 & 0.49 & \\
BRLT45 &  T1.0 $\pm$ 0.5 & 0.57 & 0.61 & 0.71 & 0.64 & 0.88 & 0.57 & 0.42 & 0.47 & SC \\
BRLT46 &  L0.5 $\pm$ 0.5 & 0.91 & 0.76 & 1.13 & 0.80 & 1.12 & 1.07 & 0.30 & 0.48 & \\
BRLT48 &  L4.5 $\pm$ 0.5 & 0.77 & 0.77 & 1.04 & 0.81 & 1.19 & 1.08 & 0.55 & 0.49 & \\
BRLT50 & T7.0 $\pm$ 0.5 & 0.16 & 0.33 & 0.84 & 0.28 & 0.25 & -0.0 & 0.06 & 0.20 & \\
BRLT51 &  L3.0 $\pm$ 1.0 & 0.80 & 0.78 & 0.86 & 0.83 & 1.06 & 0.92 & 0.51 & 0.47 & \\
BRLT52 &  L5.5 $\pm$ 0.5 & 0.75 & 0.68 & 0.94 & 0.77 & 1.05 & 0.94 & 0.57 & 0.50 & \\
BRLT56 &  L1.5 $\pm$ 1.0 & 0.91 & 0.86 & 0.93 & 0.85 & 1.01 & 0.96 & 0.44 & 0.49 & \\
BRLT57 &  L0.0 $\pm$ 1.0 & 0.97 & 0.86 & 1.08 & 0.84 & 1.09 & 0.99 & 0.37 & 0.50 & \\
BRLT58 &  L5.5 $\pm$ 1.0 & 0.76 & 0.72 & 0.84 & 0.74 & 1.05 & 1.01 & 0.41 & 0.49 & \\
BRLT60 &  L1.0 $\pm$ 1.0 & 0.83 & 0.79 & 0.99 & 0.84 & 0.99 & 0.94 & 0.41 & 0.46 & WC \\
BRLT62 &  L5.0 $\pm$ 1.0 & 0.83 & 0.76 & 0.97 & 0.86 & 1.11 & 1.04 & 0.63 & 0.49 & \\
BRLT64 &  L4.0 $\pm$ 0.5 & 0.76 & 0.73 & 0.92 & 0.83 & 1.05 & 1.01 & 0.50 & 0.48 & \\
BRLT66 &  L5.0 $\pm$ 0.5 & 0.74 & 0.69 & 0.92 & 0.79 & 1.08 & 0.95 & 0.52 & 0.48 & \\
BRLT305 &  L5.5 $\pm$ 1.0 & 0.93 & 0.79 & 0.95 & 0.91 & 0.97 & 0.93 & 0.61 & 0.47 & WC \\
BRLT306 &  L4.5 $\pm$ 1.0 & 0.81 & 0.80 & 0.95 & 0.75 & 1.01 & 0.92 & 0.39 & 0.48 & \\
BRLT307 &  L1.0 $\pm$ 0.5 & 0.92 & 0.89 & 1.17 & 0.93 & 1.07 & 0.93 & 0.47 & 0.48 & \\
BRLT311 &  T3.0 $\pm$ 0.5 & 0.42 & 0.56 & 0.78 & 0.55 & 0.70 & 0.57 & 0.29 & 0.44 & SC \\
BRLT312 &  T0.0 $\pm$ 0.5 & 0.62 & 0.67 & 0.87 & 0.66 & 0.92 & 0.90 & 0.44 & 0.42 & SC \\
BRLT313 &  L3.5 $\pm$ 0.5 & 0.82 & 0.74 & 0.94 & 0.88 & 1.04 & 0.99 & 0.53 & 0.47 & \\
BRLT314 &  L7.5 $\pm$ 0.5 & 0.72 & 0.71 & 0.88 & 0.80 & 1.05 & 0.95 & 0.69 & 0.49 & \\
BRLT315 &  L1.0 $\pm$ 1.0 & 0.87 & 0.82 & 0.92 & 0.78 & 1.04 & 1.01 & 0.35 & 0.47 & \\
BRLT316 &  L1.0 $\pm$ 0.5 & 0.87 & 0.87 & 0.91 & 0.81 & 1.13 & 0.98 & 0.35 & 0.49 & \\
BRLT317 &  L1.0 $\pm$ 1.0 & 0.79 & 0.71 & 0.92 & 0.70 & 1.09 & 0.95 & 0.31 & 0.49 & \\
BRLT318 &  L1.0 $\pm$ 0.5 & 0.83 & 0.88 & 0.97 & 0.85 & 1.13 & 0.95 & 0.42 & 0.49 & \\
BRLT320 &  L1.0 $\pm$ 0.5 & 0.89 & 0.79 & 0.90 & 0.85 & 1.10 & 0.89 & 0.42 & 0.50 & \\
BRLT321 &  T4.0 $\pm$ 0.5 & 0.37 & 0.42 & 0.57 & 0.43 & 0.60 & 0.25 & 0.24 & 0.37 & \\
BRLT322 &  L5.0 $\pm$ 0.5 & 0.71 & 0.70 & 0.90 & 0.72 & 1.07 & 0.90 & 0.55 & 0.52 & \\
BRLT323 &  T0.0 $\pm$ 1.0 & 0.76 & 0.71 & 0.87 & 0.76 & 1.04 & 1.04 & 0.42 & 0.49 & \\
BRLT325 &  T2.0 $\pm$ 1.0 & 0.53 & 0.47 & 0.63 & 0.55 & 0.89 & 0.49 & 0.27 & 0.48 & \\
BRLT328 &  L0.5 $\pm$ 1.0 & 0.89 & 0.78 & 0.97 & 0.82 & 1.09 & 1.05 & 0.36 & 0.50 & \\
\hline
\end{tabular}
\label{specinfo}
\normalsize
\end{table*}
\begin{table*}
\small
\raggedright{\bf Table~\ref{specinfo} continued.} Spectral details of observed targets.\\
\centering
\begin{tabular}{|l|c|c|c|c|c|c|c|c|c|c|}
\hline
ID& Spectral type & H$_2$O-J & H$_2$O-H & H$_2$O-K & CH$_4$-J & CH$_4$-H & CH$_4$-K & K/J & H-dip & Notes \\
\hline
BRLT330 &  L1.0 $\pm$ 1.0 & 0.81 & 0.77 & 0.89 & 0.73 & 1.20 & 0.95 & 0.35 & 0.48 & \\
BRLT331 &  L1.0 $\pm$ 2.0 & 0.84 & 0.71 & 1.06 & 0.74 & 1.05 & 0.93 & 0.30 & 0.49 & \\
BRLT332 &  L2.0 $\pm$ 2.0 & 0.87 & 0.85 & 1.00 & 0.75 & 1.10 & 0.96 & 0.34 & 0.49 &\\
BRLT333 &  T2.0 $\pm$ 0.5 & 0.52 & 0.57 & 0.77 & 0.59 & 0.98 & 0.72 & 0.38 & 0.52 & SC \\
BRLT334 &  L3.5 $\pm$ 0.5 & 0.83 & 0.74 & 0.91 & 0.82 & 1.04 & 0.95 & 0.48 & 0.48 & \\
BRLT335 &  L4.0 $\pm$ 1.0 & 0.81 & 0.74 & 0.95 & 0.79 & 0.99 & 0.92 & 0.42 & 0.47 & WC \\
BRLT338 &  L1.0 $\pm$ 1.0 & 0.88 & 0.73 & 1.01 & 0.81 & 1.09 & 1.12 & 0.34 & 0.49 & \\
BRLT343 &  L9.0 $\pm$ 1.0 & 0.69 & 0.70 & 0.88 & 0.75 & 1.05 & 0.92 & 0.60 & 0.49 & \\
BRLT344 &  T0.0 $\pm$ 1.0 & 0.54 & 0.68 & 0.96 & 0.68 & 0.94 & 0.89 & 0.54 & 0.44 & SC \\
\hline
\multicolumn{11}{l}{SC=Strong binary candidate, WC=Weak binary candidate}\\
\hline
\end{tabular}
\normalsize
\end{table*}

\subsection{Unresolved binarity}
\label{bins}
The L-T transition region is thought to be hindered with a substantial number of unresolved binaries (up to 45\%; \citealt{maxted05}), such that in order to measure a reliable formation history we must distinguish between these binaries. With the wavelength coverage of our X-shooter data we should be able to differentiate between binaries where the components are separated by more than 1 spectral type, leaving only the equal spectral-type binaries, which should not affect the overall shape of the $T_{\rm eff}$ distribution, since the equal $T_{\rm eff}$ binary fraction should be uniform over this spectral type range \citep{geissler2011}.
  
\begin{figure*}
\includegraphics[width=180mm,angle=0]{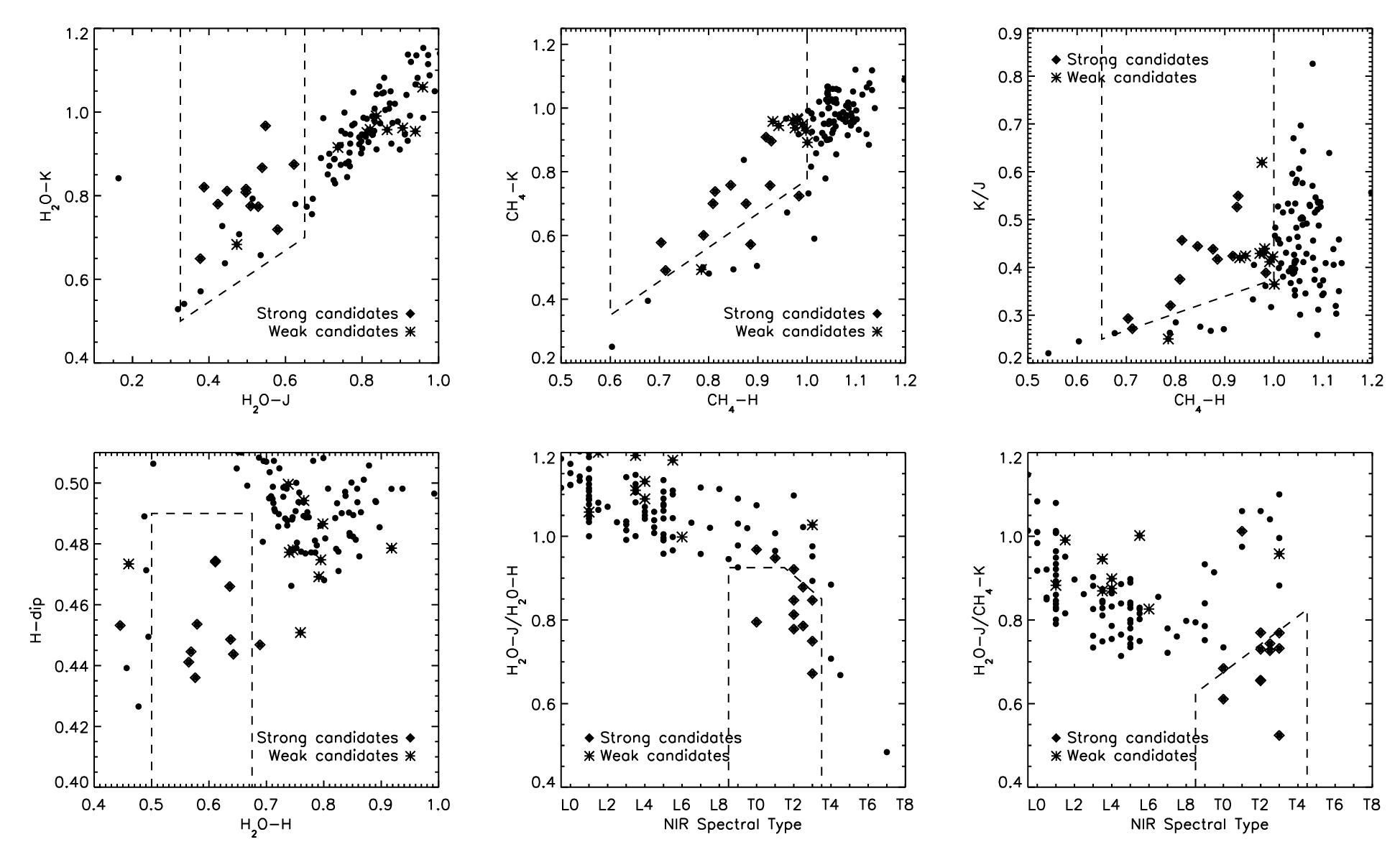}
\caption{Index-Index plots showing the selection criteria applied for unresolved binary candidate segregation. Strong and weak candidates are marked as diamonds and crosses respectively.}
\label{bins_select}
\end{figure*}

We select potential binary candidates using the spectral indices and the criteria defined in \citet{burgy10} who use index-index and index-spectral type combinations to define six different criteria to segregate possible unresolved L-T transition binaries, using known double objects as reference. These spectral indices take into account the flux ratio of the prominent molecular absorption bands of water and methane over the flux emitted in the $J$, $H$ and $K$ bands. Objects that match two criteria are called $“$weak candidates$”$ while objects that match at least three of them are called $“$strong candidates$”$. For clarity, all criteria used to classify these dwarfs are re-stated in Table.~\ref{binindex1} and ~\ref{binindex2}. We show our observed sample and the binary selection criteria applied in the different index-index and index-spectral type in Fig.~\ref{bins_select}. In this way we identified six  $“$strong$“$ candidates and six  $“$weak $“$ candidates, which are highlighted as diamonds and asterisks, respectively. 

\begin{table}
\caption{Binary indices as defined by ~\citet{burgy06a}, except for \textit{H}-dip, which is defined by ~\citet{burgy10}.}
\begin{tabular}{|l c c c|}
\hline
Index & Numerator & Denominator & Feature \\
 & Range & Range & \\
\hline
H$_2$O-\textit{J} & 1.14-1.165  & 1.26-1.285 & 1.15 $\mu$m H$_2$O \\
H$_2$O-\textit{H} & 1.48-1.52   & 1.56-1.60  & 1.4 $\mu$m H$_2$O \\
H$_2$O-\textit{K} & 1.975-1.995 & 2.08-2.10  & 1.9 $\mu$m H$_2$O \\
CH$_4$-\textit{J} & 1.315-1.34  & 1.26-1.285 & 1.32 $\mu$m CH$_4$ \\
CH$_4$-\textit{H} & 1.635-1.675 & 1.56-1.60  & 1.65 $\mu$m CH$_4$ \\
CH$_4$-\textit{K} & 2.215-2.255 & 2.08-2.12  & 2.2 $\mu$m CH$_4$ \\
\textit{K}/\textit{J}      & 2.060-2.10  & 1.25-1.29  & \textit{J-K} colour \\
\textit{H}-dip    & 1.61-1.64 & 1.56-1.59 + 1.66-1.69 & 1.65 $\mu$m CH$_4$ \\
\hline
\end{tabular}
\label{binindex1}
\end{table}

\begin{table}
\caption{Inflection points used for analysis of unresolved binarity \citet{burgy10}.}
\begin{tabular}{|l c c|}
\hline
Abscissa & Ordinate & Inflection Points \\
 & & (x,y) \\
\hline
H$_2$O-\textit{J} & H$_2$O-\textit{K} & (0.325,0.5),(0.65,0.7) \\
CH$_4$-\textit{H} & CH$_4$O-\textit{K} & (0.6,0.35),(1,0.775) \\
CH$_4$-\textit{H} & \textit{K}/\textit{J} & (0.65,0.25),(1,0.375) \\
H$_2$O-\textit{H} & \textit{H}-dip & (0.5,0.49),(0.875,0.49) \\
SpT & H$_2$O-\textit{J}/H$_2$O-\textit{H} & (L8.5,0.925),(T1.5,0.925),(T3.5,0.85) \\
SpT & H$_2$O-\textit{J}/CH$_4$-\textit{K} & (L8.5,0.625),(T4.5,0.825) \\
\hline
\end{tabular}
\label{binindex2}
\end{table}

\begin{figure*}
\includegraphics[width=180mm,angle=0]{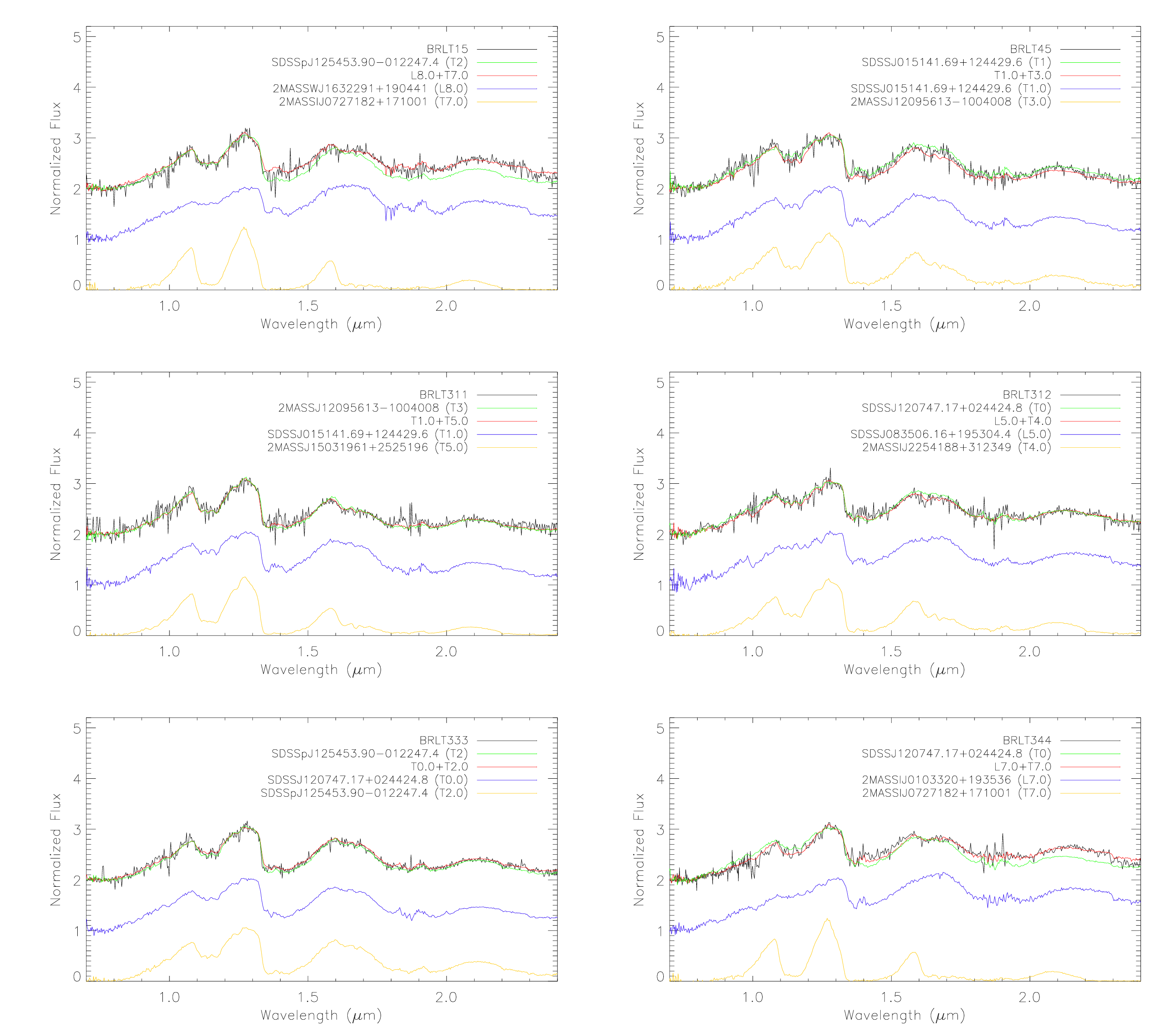}
\caption{$"$Strong$"$ unresolved binary candidates over-plotted with combined spectral templates.}
\label{binfits1}
\end{figure*}

\begin{figure*}
\includegraphics[width=180mm,angle=0]{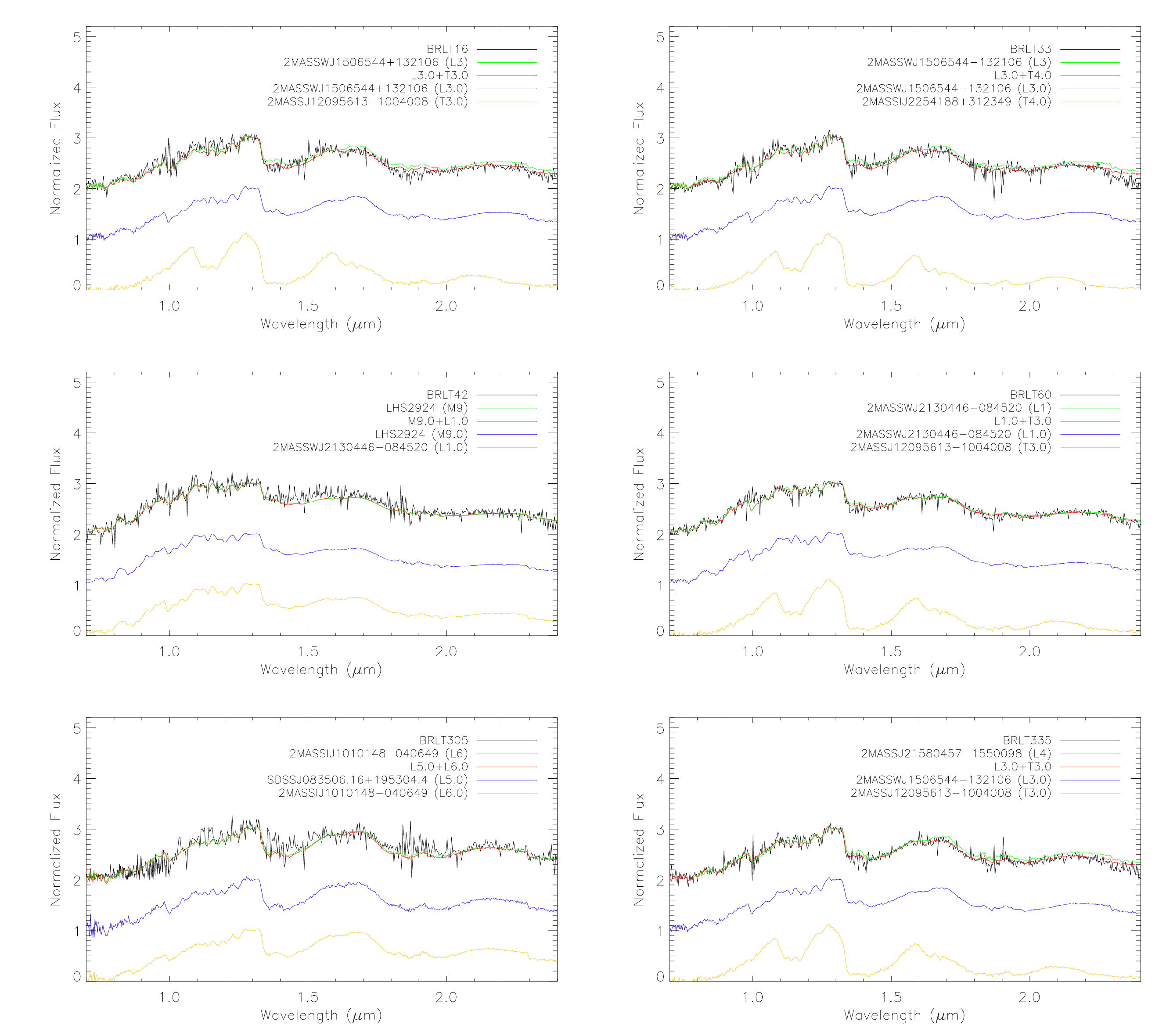}
\caption{$"$ Weak$"$ unresolved binary candidates over-plotted with combined spectral templates.}
\label{binfits2}
\end{figure*}

The 12 unresolved binary candidates were investigated further using synthetic binary template fitting. The synthetic binaries were created by taking the standard spectral type templates available in the SpeX-Prism library. These were then normalized to 1 at 1.28$\mu$m and scaled the spectra to an absolute flux level, using the absolute $J$ magnitude - spectral type relation derived in \citet{marocco10}. Finally all the templates were combined to make our own set of synthetic unresolved binaries. We then performed a chi-squared fit of our spectra to the templates. The results of the fitting are presented in Table.~\ref{bin_fitting}, with the best-fitting combined template over-plotted on the spectra of the 12 binary candidates shown in Fig.~\ref{binfits1} and ~\ref{binfits2}. For each object we list the best fit single object template and the associated chi-squared, the best fit combined template and its associated chi-squared fit value.  The results of the two fits were compared using a one-sided F test to assess their statistical significance. If the ratio of the two chi-squared fits ($\eta$) is greater than the critical value (1.15), this represents a 99$\%$ significance that the combined template fit is better than the standard template alone. The last column of Table \ref{bin_fitting} presents the results of the F test, where it can be seen that six of the 12 candidates give statistically better fits using the combined templates, such that they are the strongest binary candidates. It should be noted that the results of this fitting are not conclusive \citep{burgy10}, and the real nature of these objects must be investigated further using adaptive optics imaging or spectro-astrometry in order to confirm or not their true binary nature. 

\begin{table}
\label{bin_fitting}
\caption{Results of spectral fitting of binary candidates using combined templates.}
\footnotesize
\begin{tabular}{|l c c c|}
\hline
\normalsize Target & \normalsize Single template & \normalsize Combined template & \normalsize F-test \\
\normalsize ID & \normalsize best fit ($\chi^2$) & \normalsize best fit ($\chi^2$) & $\eta$ \\
\hline
\multicolumn{4}{|c|}{\normalsize Strong candidates} \\
\hline
BRLT15  & T2.0 (14.8341) & L8.0+T7.0 (3.6926) & 4.02 \\
BRLT45  & T1.0 (6.2583)  & T1.0+T3.0 (6.7607) & 0.92 \\
BRLT311 & T3.0 (9.3321)  & T1.0+T5.0 (7.2732) & 1.28 \\
BRLT312	& T0.0 (6.5036)  & L5.0+T4.0 (5.3935) & 1.20 \\
BRLT333	& T2.0 (3.5947)  & T0.0+T2.0 (3.7963) & 0.95 \\
BRLT344 & T0.0 (5.8038)  & L7.0+T7.0 (3.5877) & 1.62 \\
\hline
\multicolumn{4}{|c|}{\normalsize Weak candidates} \\
\hline
BRLT16	& L3.5 (3.7444) & L3.0+T3.0 (3.6042) & 1.04 \\
BRLT33	& L3.5 (6.0794) & L3.0+T4.0 (4.4195) & 1.37 \\
BRLT42 	& M9.0 (5.2356) & M9.0+L1.0 (3.3914) & 1.54 \\
BRLT60 	& L1.0 (2.2994) & L1.0+T3.0 (2.1760) & 1.06 \\
BRLT305 & L5.5 (7.0290) & L5.0+L6.0 (8.9120) & 0.79 \\
BRLT335 & L4.0 (4.6314) & L3.0+T3.0 (4.2767) & 1.08 \\
\hline
\end{tabular}
\end{table}

\section{Constraining the Galactic brown dwarf formation history}
\label{fh}

In order to compare the findings from our first sub-sample of 76  mid L to mid T dwarfs that occupy a complete area of 495 sq degrees of sky (RA=22 to 4~hr and DEC=-2 to 16~deg) down to a magnitude limit of J$=$18.1, with the results from simulations of differing birth rates, we constructed a space density vs  spectral type/$T_{\rm eff}$ histogram. We firstly calculated the space density of our targets by converting their spectral types into effective temperatures using the $T_{\rm eff} -$ NIR spectral type relation presented in \citet[][ equation 3]{stephens2009}. Using a bin sizes of 1700-1450K, 1150-1300K and 1300-1450K (corresponding to bins in spectral type of L4-L6, L7-T0 and T1-T4), we calculated for each bin the maximum distance at which an object could have been selected using the $M_J -$ NIR spectral type relation from \citet{marocco10}. With this distance limit we then calculated the volume sampled by each $T_{\rm eff}$ bin. 
The derived space densities were then corrected for Malmquist and Eddington bias following the approach described in \citet{pinfield08}. The Eddington bias is caused by the photometric uncertainties on the magnitudes of objects near our cut (i.e. J $<$ 18.1). However, since the magnitude cut we impose is bright (it corresponds to a $\sim 12 \sigma$ detection in the UKIDSS LAS), the uncertainties at the J=18.1 limit are typically less than $\sigma$=0.05 and therefore the Eddington bias correction is less than 1$\%$. This is negligible compared to the other sources of uncertainty. We estimated the Malmquist bias correction considering the mean scatter of the sample of known L and T dwarfs around the adopted $M_J -$ NIR spectral type relation. This represents an increase in the volume sampled of 22$\%$. 

\subsection{Completeness}
\label{comp}
In order to calculate the completeness of our sample we compared it to a control sample of known L and T dwarfs taken from DwarfArchives.org, for a magnitude limit of $J \le 16$, removing any objects that are known to be members of unresolved binary systems. The control sample was cross matched with the UKIDSS LAS and SDSS in order to obtain photometry on the same colour system as our selection criteria. We use DR9 of the UKIDSS LAS and DR8 of the SDSS in order to get a sufficiently large control sample. We imposed the same set of colour cuts, as described in $\S$\ref{sample} to reveal the level of completeness of our sample selection. We retain all of the L4 dwarfs from the control sample, but only some of the L0-L3 dwarfs when imposing our colour cuts, as such we find that our sample selection is complete for L4 spectral types and later. 

We also consider how many objects we may lose from our selection due to photometric scattering of colours. We calculate that for 
L0-L3 types we would expect to lose 3.7 dwarfs, this corresponds to a completeness level of 85$\%$ for L0-L3 dwarfs due to photometric scatter, although we note that we are significantly incomplete from our colour selection for this spectral range. For L4-L6 dwarfs we would expect to lose 2.33 dwarfs, being 88$\%$ complete. The L7-T0 range would lose 0.55 dwarfs, corresponding to a 94$\%$ completeness; and for T1-T4  we would expect to lose 0.05 dwarfs, corresponding to a completeness of 99$\%$. 

We combine the completeness for colour selection, colour scatter as well as the Malmquist and Eddington biases to infer a 90$\%$ completeness for the $T_{\rm eff}$ =1450-1700K and 97$\%$ for the $T_{\rm eff}$=1300-1450K and 97$\%$ for the $T_{\rm eff}$=1150-1300K range in our sample.

\subsection{Correction for unresolved binarity}
\label{bincorrection}
We also corrected our results for the presence of binaries by firstly
removing objects identified as possible binaries ($\S$\ref{bins}) for
which our spectral deconvolution gives a statistically better fit. We then
applied a further correction to take into account the presence of equal
spectral type binaries, which would fall beyond our J $<$ 18.1 limit if they
were single objects, using the definition of ``observed binary fraction''
given by \citet{burgy03}, such that:
 
\begin{equation}
\frac{N_B}{N_m} = \frac{\gamma + 1}{BF - 1}
\end{equation}
 
where $N_B$ and $N_m$ are the observed binaries and the total number
of objects respectively, BF is the ``true'' binary fraction, and
$\gamma$ is the fractional increase in volume due to inclusion of
binaries in the sample. The number of binaries that fall within our
magnitude limit ($N_D$) is:
 
\begin{equation}
\frac{N_D}{N_B} = \frac{\gamma -1}{\gamma}
\end{equation}
 
Therefore the fraction of objects to be excluded from our sample ($f_{excl}$) is
\begin{equation}
f_{excl} = \frac{N_B}{N_m} \frac{N_D}{N_B} = \frac{\gamma -1}{\gamma + 1/BF - 1}
\end{equation} 

For equal spectral type binaries $\gamma$ = 2$\sqrt{2}$. For the
``true'' binary fraction we assumed two extremes values taken from the
literature. Following \citet{burningham10a}, for the lower limit
we considered the values obtained by \citet{burgy03} who
estimated a BF of 5-24$\%$ using AO imaging; for the upper limit we
used the values given by \citet{maxted05}, who estimated a
BF of 32-45$\%$ via radial velocity monitoring.

\subsection{Space densities}
\label{density}
We combine our calculated space densities, taking into account our completeness and contribution from unresolved binaries, with those from \citet{burningham10a} for late-type T dwarfs. We choose to use the \citet{burningham10a} space densities in favour of others available in the literature (e.g. \citealt{cruz07}, \citealt{metchev08}, \citealt{reyle10}, \citealt{kp12}) as this is the only space density that can be used in direct comparison to our own sample, as we use the same M$_{J}$-spectral type relations as well as apply the same binary fractions. It is also probes down to a magnitude limit comparable to our sample. The \citet{cruz07} space densities only probe out to the 2MASS limit ($\sim J=16$) and covers only M9-L8 dwarfs, and not the full range of spectral types we are probing. The \citet{metchev08} sample provide space densities obtained from 2MASS, combined with SDSS, and as such does not probe as deeply as our sample, or completely into the late T dwarf spectral type as that of \citet{burningham10a} .  The \citet{reyle10} derived space densities could also be comparable but for their very late T type densities (T6-T8), they only have a handful of objects and do not make any correction for binarity. As such the \citet{burningham10a} space densities act as a complementary extension to our mid L-mid T derived space densities. Our calculated space densities are shown in Table.\ref{spacedensities}, with comparison to the above mentioned published space densities for L and T dwarfs.

\begin{figure*}
\includegraphics[width=180mm,angle=0]{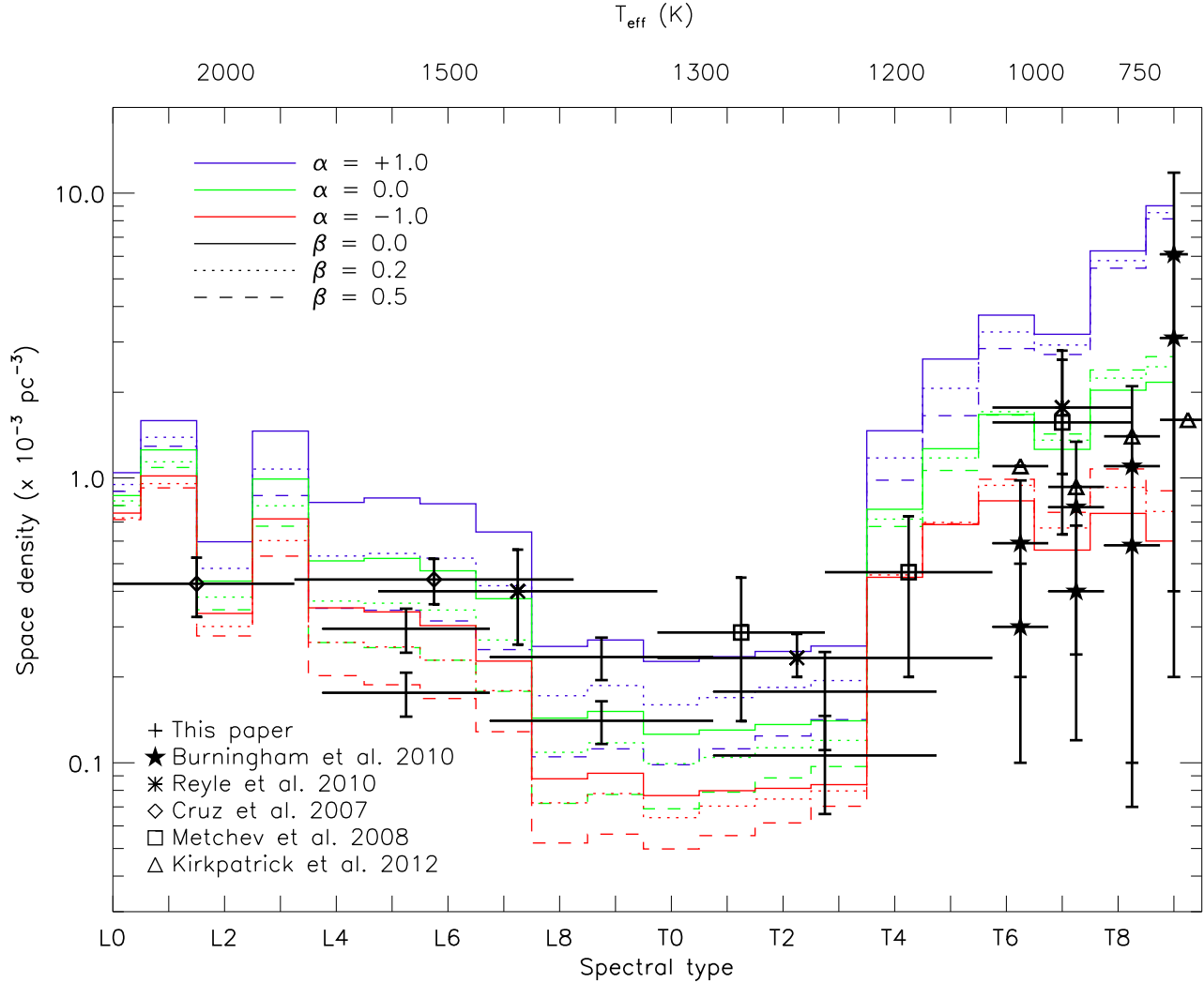}
\caption{Our sub-sample of mid L-mid T dwarfs are overplotted with simulations from \citet{deacon2006} with $\alpha$ = +1.0, 0.0, -1.0 and $\beta$ = 0.0, 0.2, 0.5 (where $\psi(M)~\propto~M^{-\alpha}$). Upper most points of the bins represent a binary fraction of 5$\%$ and the lower points are for a binary fraction of 45$\%$.}
\label{spthistogram}
\end{figure*}

\begin{figure*}
\includegraphics[width=180mm,angle=0]{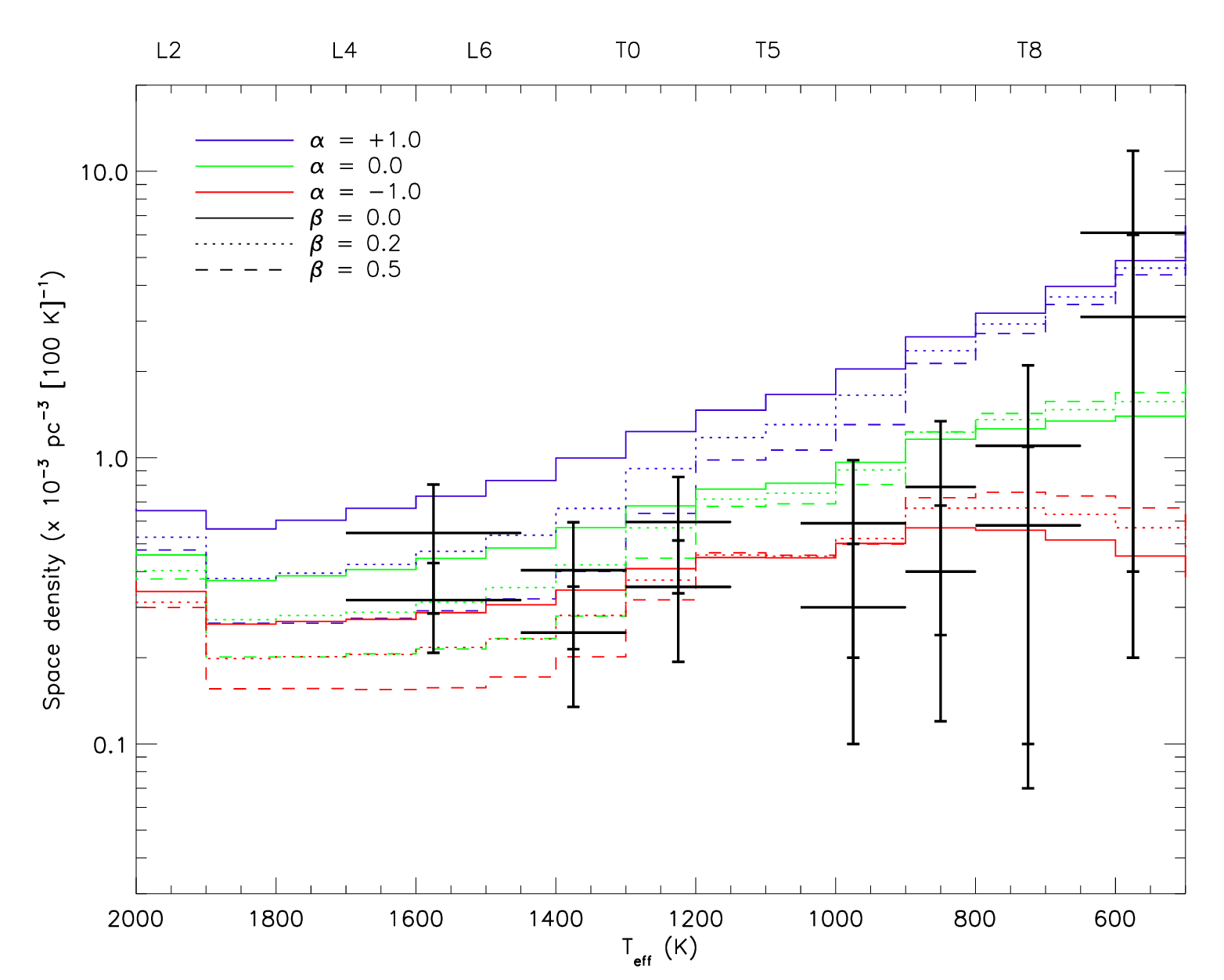}
\caption{Our sub-sample of mid L-mid T dwarfs are overplotted with simulations from \citet{deacon2006} with $\alpha$ = +1.0, 0.0, -1.0 and $\beta$ = 0.0, 0.2, 0.5 (where $\psi(M)~\propto~M^{-\alpha}$). With bins of 1700-1450K, 1450-1300K and 1300-1150K, and 500-1100K for late type T dwarfs as taken from \citet{burningham10a}. Upper most points of the bins represent a binary fraction of 5$\%$ and the lower points are for a binary fraction of 45$\%$.}
\label{histogram}
\end{figure*}

Our space densities are also shown as a function of both spectral type in Fig.~\ref{spthistogram}, and as a function of $T_{\rm eff}$ in Fig.~\ref{histogram}, where we show two points for each of our $T_{\rm eff}$/spectral type bin, the upper most points represent the density corresponding to a binary fraction of just 5$\%$, while the lower points are those for a binary fraction of 45$\%$. Also overplotted are space densities published for L and T dwarfs. In addition we show (overplotted) the results of numerical simulations computed assuming different IMF and birth rates. Details of the simulations are presented in \citet{deacon2006} and are briefly summarized here. 

We assume an exponential IMF in the form:

\begin{equation}
\psi(M)~\propto~M^{-\alpha}~(pc^{-3} \Msun^{-1}).
\label{imf}
\end{equation}

where $\Psi$ is the number of objects per unit volume in a given mass interval.
We also assumed an exponential birth rate of the form:

\begin{equation}
b(t) \propto e^{-\beta t}
\label{br}
\end{equation}

where $t$ is in Gyr and $\beta$ is the inverse of the scale time $\tau$ (in Gyr, since the galaxy was formed). Each simulated object was assigned an age based on the birth rate and a mass based on the IMF, giving a final creation function $C$ given by the equation:

\begin{equation}
C(M,t) = \Psi(M) \frac{b(t)}{T_{\rm G}}
\end{equation}

where $T_{\rm G}$ is the age of the Galaxy. $C$ is therefore the number of objects created per unit time per unit mass.

The evolution of each object and its parameters (i.e. $T_{\rm eff}$ and absolute magnitudes) were calculated using the evolutionary models from \citet{baraffe2003}. We note that any any model dependent systematics would be introduced, but that these should not effect the overall trend. The number densities obtained for each temperature bin were finally normalized to 0.0024 pc$^{-3}$ in the 0.1$-$0.09 M$_\odot$ mass range, according to \citet{deacon08}. 

We consider five different values of $\beta$ (-0.2,-0.1,0.0,+0.1,+0.2 corresponding to $\tau$=-5,-10,$\infty$,+10,+5 Gyr respectively) and five values of $\alpha$ (0.0,-0.5,-1.0-,1.5,-2.0). The results obtained for $\alpha$ = 0.0,-1.0,-2.0 and $\beta$ = 0.0,0.2,0.5 are shown in Figure ~\ref{histogram}, where different colours represent different values of $\alpha$ and different line styles represent different values of $\beta$.

A first look at our measured mass function shows that the more extreme birth rates (e.g. a halo type form) can likely be excluded as the true form of the birth rate. Our space densities are in general agreement with the previous Bayesian analysis of 2MASS L and T dwarfs by \citet{allen05}, and do not differ drastically (within our uncertainties) from those mass functions previously measured and discussed earlier (i.e. \citealt{cruz07}; \citealt{metchev08}; \citealt{reyle10}; \citealt{burningham10a}; \citealt{kp12}). The differences between our derived densities and those previously published can be accounted for in large due to the use of different M$_{J}$-SpT conversions by the different groups.

Our space densities do however suggestive of a preference for an $\alpha < -1.0$, which is consistent with the studies of late T dwarfs alone from \citet{pinfield08} and \citet{burningham10a}, where they conservatively find $\alpha <0$. Indeed all three samples are indicative of an $-1<\alpha < 0$. It can be seen however that it is not currently possible to place robust constraints on the birth rate with this sub-sample. One of the largest sources of uncertainties is the binary fraction. This could be resolved with the follow-up of our unresolved binary candidates, by either adaptive optics imaging or radial velocity. In addition improved numbers from the completion of follow-up of our full sample would allow us to place better constraints on the birth rate and would allow us to rule out at least $\beta =0$ or $\beta =0.5$ as the uncertainties are reduced. With spectroscopic follow-up of a full sample the errors could be reduced by up to $50\%$.

\begin{table*}
\caption{Calculated space densities.} 
\begin{tabular}{|l c c c c c|}
\hline
Reference  & Sp. Type & Space density & Malmquist and Eddington & Binary & Space density \\
 & range & (no correction) & bias correction (\%) & correction (\%) & ($\times 10^{-3}$ pc$^{-3}$) \\
 & & ($\times 10^{-3}$ pc$^{-3}$) & & & \\
\hline
This paper & L4 - L6.5 & 0.345 $\pm$ 0.061 & 22 & 3-45 & $0.176 \pm 0.031 - 0.295 \pm 0.052$ \\
 & L7 - T0.5 & 0.293 $\pm$ 0.050 & 22 & 3-45 & $0.140 \pm 0.024 - 0.235 \pm 0.040$ \\
 & T1 - T4.5 & 0.235 $\pm$ 0.088 & 22 & 3-45 & $0.106 \pm 0.040 - 0.178 \pm 0.067$ \\
\hline
Burningham et al. (2010a) & T6 - T6.5 & - & 12-16 & 3-45 & $0.30 \pm 0.2 - 0.59 \pm 0.39$ \\
 & T7 - T7.5 & - & 12-16 & 3-45 & $0.40 \pm 0.28 - 0.79 \pm 0.55$ \\
 & T8 - T8.5 & - & 12-16 & 3-45 & $0.58 \pm 051 - 1.1 \pm 1.0$ \\
 & T9        & - & 12-16 & 3-45 & $3.1 \pm 2.9 - 6.1 \pm 5.7$ \\
\hline
Reyle et al. (2010) & L5 - L9.5 & - & - & - & $2.0^{+0.8}_{-0.7}$ \\
 & T0 - T5.5 & - & - & - & $1.4^{+0.3}_{-0.2}$ \\
 & T6 - T8   & - & - & - & $5.3^{+3.1}_{-2.2}$ \\
\hline
Cruz et al. (2007) & M7 - M9.5 & - & - & - & $4.9 \pm 0.6$ \\
 & L0 - L3 & - & - & - & $1.7 \pm 0.4$ \\
 & L3.5 - L8 & - & - & - & $2.2 \pm 0.4$ \\
\hline
Metchev et al. (2008) & T0 - T2.5 & - & - & $\sim$50 & $0.86^{+0.48}_{-0.44}$ \\
 & T3 - T5.5 & - & - & $\sim$22 & $1.4^{+0.8}_{-0.8}$ \\
 & T6 - T8   & - & - & $\sim$14 & $4.7^{+3.1}_{-2.8}$ \\
\hline
Kirkpatrick et al. (2012) & T6 - T6.5 & - & - & 30 & $1.1$ \\
 & T7 - T7.5 & - & - & 30 & $0.93$ \\
 & T8 - T8.5 & - & - & 30 & $1.4$ \\
 & T9 - T9.5 & - & - & 30 & $1.6$ \\
 & Y0 - Y0.5 & - & - & 30 & $1.9$ \\
\hline
\end{tabular}
\label{spacedensities}
\end{table*}

\section{Summary}
\label{concs}
We have identified 63 new brown dwarfs that lie in the L-T transition region, including the identification of 12 possible unresolved binary systems. In order to confirm these as real binary systems additional adaptive optics/ radial velocity measurements are planned. We converted our 76 strong new sample onto a histogram of spatial density vs $T_{\rm eff}$/SpT and compared them with simulations of differing birth rates from \citet{deacon2006}. This showed that the mid L-mid T range are in general agreement (within the errors) with the indications from the late T dwarfs alone from \citet{pinfield08} and  \citet{burningham10a}, such that $\alpha <0$. Indeed both our L-T transition sample and the late T dwarf samples are suggestive of $-1< \alpha < 0$. A better constraint on the binary fraction, as well as a larger sample size however, is required before we can place robust constraints on the form of the brown dwarf birth rate, other than to suggest that a halo form of the birth rate is extremely unlikely.

\section*{Acknowledgments}
ADJ is supported by a FONDECYT postdoctorado fellowship under project
number 3100098. ADJ is also partially supported by the Joint Committee ESO-Government of Chile. JJ is also supported by a FONDECYT postoctorado fellowship (project number 3110004). ADJ, JJ and MTR would also like to acknowledge the support of the grant from CONICYT and the partial support from Center for Astrophysics FONDAP
and Proyecto Basal PB06 (CATA). ADJ and JJ are also partially supported by the Joint Committee ESO-Government of Chile. JG is supported by RoPACS, a Marie Curie Initial Training Network funded by the European Commission's Seventh Framework Programme. SC is supported by a Marie Curie Intra-European Fellowship within the 7th European Community Framework Programme. This research has benefited from the SpeX Prism Spectral Libraries, maintained by Adam Burgasser at http://pono.ucsd.edu/~adam/browndwarfs/spexprism.

\appendix

\bsp

\label{lastpage}

\end{document}